\pdfoutput=1

\documentclass[twocolumn,
  showkeys,amssymb,amsmath,aps,nofootinbib,floatfix,superscriptaddress]{revtex4-1}

\usepackage{epsfig}
\usepackage{hyperref}

\begin{document} 

\title{Transverse momentum fluctuations in ultrarelativistic Pb+Pb and p+Pb collisions with wounded quarks}

\author{Piotr Bo\.zek}
\email{Piotr.Bozek@fis.agh.edu.pl}
\affiliation{AGH University of Science and Technology, Faculty of Physics and
Applied Computer Science, al. Mickiewicza 30, 30-059 Cracow, Poland}

\author{Wojciech Broniowski}
\email{Wojciech.Broniowski@ifj.edu.pl}
\affiliation{The H. Niewodnicza\'nski Institute of Nuclear Physics, Polish Academy of Sciences, 31-342 Cracow, Poland}
\affiliation{Institute of Physics, Jan Kochanowski University, 25-406 Kielce, Poland}

\begin{abstract}
We analyze the phenomenon of size-flow transmutation in ultrarelativistic 
nuclear collisions in a model where the initial size fluctuations are driven by 
the wounded quarks and the collectivity is provided by  viscous hydrodynamics. 
It is found that the model properly reproduces the data for the transverse momentum fluctuations 
measured for Pb+Pb collisions at $\sqrt{s_{NN}}=2.76$~TeV by the ALICE Collaboration.
The agreement holds for a remarkably wide range of centralities, from 0-5\% up to 70-80\%, 
and displays a departure from a simple scaling with $(dN_{\rm ch}/d\eta)^{1/2}$ in the form seen in the data.
The overall agreement in the model with wounded quarks is significantly better than 
with nucleon participants. This feature joins the previously-found wounded quark multiplicity scaling
in the argumentation in favor of subnucleonic degrees of freedom in the early dynamics. 
We also examine in detail the correlations between measures of the 
initial size and final average transverse momentum of hadrons.  
Predictions are made for the transverse momentum fluctuations in p+Pb collisions at $\sqrt{s_{NN}}=5.02$~TeV. 
\end{abstract}

\pacs{25.75.-q, 25.75Gz, 25.75.Ld}

\date{30 January 2017}

\keywords{ultra-relativistic nuclear collisions, transverse momentum fluctuations, size-flow transmutation, wounded quarks}

\maketitle

\section{Introduction}

Collectivity of the intermediate evolution of fireballs created in ultra-relativistic nuclear collision is by now a well accepted fact,
allowing, in particular, for a qualitative and quantitative understanding of harmonic flow phenomena as due to the initial transverse shape which fluctuates 
event by event. It is somewhat less commonly 
known that the same underlying physical effects (initial fluctuations and collectivity) jointly lead to sizable event-by-event 
transverse momentum fluctuations, one of the basic observables studied from the outset of the relativistic 
collisions program. The pertinent {\em size-flow transmutation} effect was brought up for the first time in Ref.~\cite{Broniowski:2009fm} and further elaborated in 
Ref.~\cite{Bozek:2012fw}, where we presented a detailed study 
comparing a 3+1-dimensional (3+1D) viscous hydrodynamic simulations to the data from Relativistic Heavy-Ion Collider (RHIC). 

In this paper we extend the analysis of Ref.~\cite{Bozek:2012fw} by passing to subnucleonic degrees of freedom, namely, the 
wounded quarks~\cite{Bialas:1977en,*Bialas:1977xp,*Bialas:1978ze,Anisovich:1977av}, in modeling of the initial stage. We show 
that the approach leads to a surprisingly accurate description of the recent data~\cite{Abelev:2014ckr} from the Large Hadron Collider (LHC) 
for Pb+Pb collisions at \mbox{$\sqrt{s_{NN}}=2.76$~TeV}, holding 
in a very wide range of the collision centralities, from 0-5\% to 70-80\%. Passing to subnucleonic components is physically desirable 
for other reasons, as it allows for a natural scaling 
of the multiplicity of produced hadrons on the number of (subnucleonic) participants (for a compilation of results 
see, e.g., ~\cite{Lacey:2016hqy,Bozek:2016kpf}). 
Importantly, the results for the shape eccentricities are 
similar with wounded quarks~\cite{Bozek:2016kpf} 
to models based on nucleon participants, whereby the successful phenomenology of the harmonic flow obtained with 
wounded nucleon initial conditions is maintained.

Since the transverse momentum fluctuations reveal
relevant details of the early dynamics of the  system formed in ultra-relativistic nuclear collisions, they have been intensely investigated
both theoretically
\cite{Gazdzicki:1992ri,Stodolsky:1995ds,Shuryak:1997yj,Mrowczynski:1997kz,Liu:1998xf,%
Voloshin:1999yf,Baym:1999up,Voloshin:2001ei,Korus:2001au,Gavin:2003cb,DiasdeDeus:2003ei,Voloshin:2004th,Mrowczynski:2004cg,AbdelAziz:2005wc,%
Broniowski:2005ae,Prindle:2006zz,Gavin:2006xd,Sharma:2008qr,Mrowczynski:2009wk,Hama:2009pk,Trainor:2015swa,Liu:2016apq}
and experimentally
\cite{Adams:2003uw,Adamova:2003pz,Adler:2003xq,Anticic:2003fd,Adams:2004gp,Adams:2005ka,%
Adams:2005aw,Adams:2005ka,Adams:2006sg,
Agakishiev:2011fs,Abelev:2014ckr}. 

From a broader perspective, the analysis presented in this paper contributes to the discussion of the nature of the initial stage, its degrees of freedom and 
fluctuations; the results are also to some extent sensitive to properties of the intermediate evolution (hydrodynamics, transport). Comparisons
to present and future data, made jointly with other observables, may help to resolve the issue. We also make predictions for 
the transverse momentum fluctuations in p+Pb collisions at $\sqrt{s_{NN}}=5.02$~TeV, which can be tested in future data analyses.

\section{Basic ingredients \label{sec:basic}}

\subsection{Size-flow transmutation}

\begin{figure}[tb]
\begin{center}
\includegraphics[width=0.3 \textwidth]{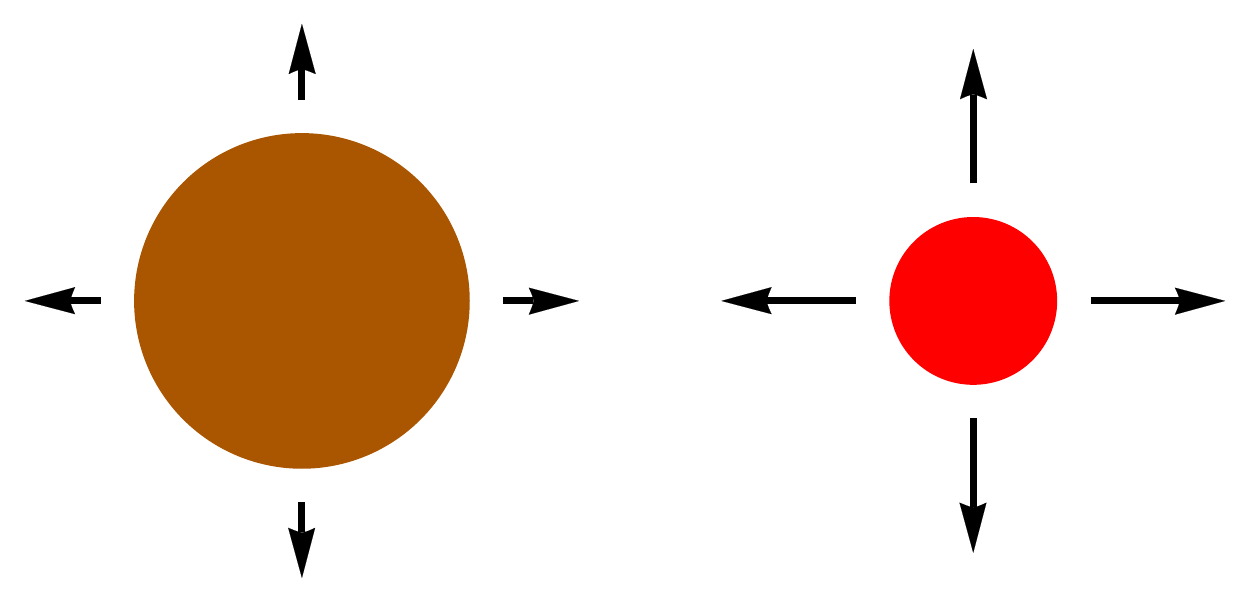} 
\end{center}
\vspace{-5mm}
\caption{A cartoon view of the size-flow transmutation effect. If two fireballs of equal entropy differ in size, a smaller one will 
lead via collective evolution to a stronger radial flow~\cite{Broniowski:2009fm}. \label{fig:trans}}
\end{figure}

We begin with a reminder of the size-flow transmutation phenomenon~\cite{Broniowski:2009fm,Bozek:2012fw}. When two nuclei collide, the number 
of participants and the transverse size of the fireball fluctuate. Even when we take a subsample with exactly the 
same number of participants, the size is slightly (a few percent) different from event to event. The amount of these fluctuations depends on a specific model 
of the nucleon structure and elementary collisions, but the effect persist as a generic phenomenon. If two fireballs created with the 
same number of participants (thus having nearly equal entropy) have different size, then the smaller one will lead to faster collective expansion
(cf. Fig.~\ref{fig:trans}). In hydrodynamics this 
is caused by a larger radial gradient of the pressure, whereas in transport models by a higher collision rate of partons. As a result, the smaller 
system leads to a larger radial flow, and consequently, a larger average transverse momentum in the event, denoted as $\langle p_T \rangle$. 
Thus, on these general grounds, we expect a strong negative correlation between the initial fireball size and $\langle p_T \rangle$.

\subsection{Wounded quarks}  

The concept of wounded quarks in the Glauber-motivated approach~\cite{Glauber:1959aa,Czyz:1969jg} to inelastic production 
in the early phase of the collision was developed in~\cite{Bialas:1977en,*Bialas:1977xp,*Bialas:1978ze,Anisovich:1977av} shortly 
after the proposal of wounded nucleons~\cite{Bialas:1976ed} (for a review see~\cite{Bialas:2008zza}). 
Over the years it has been 
argued~\cite{Eremin:2003qn,KumarNetrakanti:2004ym,Agakishiev:2011eq,Adler:2013aqf,Loizides:2014vua,Adare:2015bua,Lacey:2016hqy,Bozek:2016kpf,%
Zheng:2016nxx,Mitchell:2016jio,Loizides:2016djv} 
that the wounded quarks lead to a more natural description of the 
multiplicity of produced hadrons, with the simple scaling 
\begin{eqnarray}
\frac{dN_{\rm ch}}{d\eta} \sim Q_{\rm W}. \label{eq:wq}
\end{eqnarray}
where $Q_{\rm W}$ denotes the number of wounded quarks. 
The scaling holds to a very reasonable accuracy for a variety of reactions, including p+p collisions, and centralities~\cite{Bozek:2016kpf}, with 
the proportionality constant dependent only on $\sqrt{s_{NN}}$. The quality of scaling improves with increasing collision energy. 
The key role of the subnucleonic constituents such as the wounded quarks lies in enhanced combinatorics. 
We note that intermediate combinatorics of the quark-diquark model~\cite{Bialas:2006kw,*Bialas:2007eg} also leads to a correct description of the 
RHIC data as well as the proton-proton elastic scattering amplitude at the energies of the CERN Intersecting Storage Rings (ISR). 

Our previous study~\cite{Bozek:2012fw} involved the  wounded nucleon model, amended with binary collisions~\cite{Kharzeev:2000ph} (the so-called {\em mixed} model),
\begin{eqnarray}
\frac{dN_{\rm ch}}{d\eta} \sim \frac{1-a}{2}N_{\rm W} + a N_{\rm bin}, \label{eq:wn}
\end{eqnarray}
where $N_{\rm W}$ and  $N_{\rm bin}$ denote the number of wounded nucleons the number of binary collisions, respectively. 
The binary component is crucial for the description of the experimental data~\cite{Back:2001xy}.
With a suitable choice of the mixing parameter $a$ (at the LHC energies $a\simeq 0.15$) the model (\ref{eq:wn}) is capable  
of explaining the multiplicity distribution at RHIC and the LHC in a hydrodynamic approach~\cite{Bozek:2012qs}.
However, importantly for the message of the present work, 
the initial conditions for hydrodynamics generated with the mixed model lead to sizably (by about 50\%) too large transverse momentum fluctuations 
for the most central collisions~\cite{Bozek:2012fw}.

We remark that the wounded quarks may be regarded in more general terms of partonic degrees of freedom and generalized to more than three
constituents~\cite{Loizides:2016djv,Bozek:2016kpf}. The active constituents (those emitting to the mid-rapidity region) may also be interpreted 
in terms of partonic hot-spots (see, e.g.,~\cite{Albacete:2016pmp,Kovner:2002xa}).

\subsection{Measures of the initial size}

The initial transverse profiles of the fireball are obtained with the implementation 
of the wounded quark model in {\tt GLISSANDO}~\cite{Broniowski:2007nz,Rybczynski:2013yba},
as described in detail in Ref.~\cite{Bozek:2016kpf}. Here we only stress that care has been taken to 
properly distribute quarks within the nucleon, as well as to reproduce the nucleon-nucleon inelasticity profile and the inelasitic NN 
cross section, which are accessible experimentally.
The result of {\tt GLISSANDO} is the transverse distribution of point-like sources, which then are smeared with a Gaussian of width of 0.3~fm. 
Such a smearing is physically motivated and must always be done in applications of the Glauber modeling of the initial conditions for use in 
hydrodynamics. 

Denoting thus obtained initial transverse entropy profile in event $k$ as $s_k(x,y)$, we may define several measures of its transverse 
extent. The simplest one is the r.m.s. radius of a single event,
\begin{eqnarray}
\langle r^2 \rangle &=& \frac{\int dx\,dy\, s_k(x,y) (x^2+y^2)}{\int dx\,dy\, s_k(x,y)}, \label{eq:rd} \\
\langle r \rangle_k &\equiv& \sqrt{\langle r^2 \rangle}.  \nonumber
\end{eqnarray}
Averaging over $N_{\rm ev}$ events is denoted with another pair of brackets, for instance the event-averaged transverse size is
denoted as
\begin{eqnarray}
\langle \langle r \rangle \rangle = \frac{1}{N_{\rm ev}} \sum_{k=1}^{N_{\rm ev}} \langle r \rangle_k.
\end{eqnarray}

For sources with a large azimuthal deformation, a definition of the size parameter more appropriate for large asymmetries has been proposed
in Ref.~\cite{Bhalerao:2005mm}:
\begin{equation}
\frac{1}{\bar{R}}=\sqrt{\frac{1}{\sigma_x^2}+\frac{1}{\sigma_y^2}} \ ,
\label{eq:barr}
\end{equation}
where $\sigma_{x,y}$ denote the widths of the fireball density along its principal axes.

\subsection{3+1D viscous hydrodynamics and statistical hadronization}

In our study we use the (3+1)-D event-by-event viscous hydrodynamics~\cite{Schenke:2010rr}.
The details of the approach have been presented in Ref.~\cite{Bozek:2012fw}.  As previously, we use constant 
shear viscosity to entropy density ($s$) ratio
$\eta/s=0.08$,  constant 
bulk viscosity to $s$ ratio $\zeta/s=0.04$ (present only in the hadronic phase), whereas the 
corresponding relaxation times are $\tau_\pi={3\eta}/{(T s)}$ and
$\tau_\Pi=\tau_\pi$. The hydrodynamic evolution is started at the time $\tau_0=0.6$~fm/c.
The equation of state interpolates between the lattice-QCD results 
at high temperatures  \cite{Borsanyi:2010cj} and a hadron gas  at low temperatures~\cite{Chojnacki:2007jc,Bozek:2011ua}.

To carry out the statistical Cooper-Frye~\cite{Cooper:1974mv} hadronization at the freeze-out temperature $T_f=150$~MeV 
we run {\tt THERMINATOR}~\cite{Kisiel:2005hn,Chojnacki:2011hb}, which includes resonance decays of all hadrons listed in 
the Particle Data Tables. That way we simulate events in a close resemblance to the experiment, with the kinematic cuts as 
in the pertinent ALICE analysis of Ref.~\cite{Abelev:2014ckr}: $|\eta|<0.8$ for the pseudorapidity and $0.15 < p_T < 2$~GeV for the transverse 
momentum of the registered hadrons.

Event-by-event hydrodynamic simulations with fluctuating initial conditions have been performed for 
perfect fluid and for the viscous case, 
focusing on collective flow (for reviews, see~\cite{Gale:2013da,Heinz:2013th,Jeon:2015dfa}).

\subsection{Measure of the transverse momentum fluctuations}

Since the events consist of a large but finite number of hadrons, the $p_T$ fluctuations
involve a trivial component coming from sampling with a limited multiplicity. 
Even if the conditions at freeze-out were exactly the same in each event, we would get 
this spurious effect. Many statistical measures  have been designed to get rid of the 
trivial fluctuations. 
In our study we use a measure proposed by the STAR Collaboration~\cite{Adams:2005ka}: 
\begin{eqnarray}
\langle \Delta p_T \Delta p_T \rangle \equiv \frac{1}{N_{\rm ev}} \sum_{k=1}^{N_{\rm ev}} \frac{C_k}{N_k(N_k-1)},
\end{eqnarray}
with $N_k$ denoting the multiplicity in event $k$ and
\begin{eqnarray}
C_k=\sum_{i=1}^{N_k} \sum_{j=1,j\neq i}^{N_k} (p_{Ti}-\langle \langle p_{T} \rangle \rangle) 
(p_{Tj}-\langle \langle p_T \rangle \rangle), 
\label{star}
\end{eqnarray}
where
\begin{eqnarray}
\langle \langle p_T \rangle \rangle = \frac{1}{N_{\rm ev}} \sum_{k=1}^{N_{\rm ev}} \langle p_T \rangle_k.
\end{eqnarray}
In Ref.~\cite{Bozek:2012fw} we have shown that the measure may be more conveniently rewritten as
\begin{eqnarray}
\hspace{-7mm} \langle \Delta p_T \Delta p_T \rangle = 
\frac{N_{\rm ev}-1}{N_{\rm ev}} {\rm var}(\langle p_T \rangle)\! -\! 
\frac{1}{N_{\rm ev}} \sum_{k=1}^{N_{\rm ev}} \frac{{\rm var}_k ( p_T )}{N_k},  \label{eq:my}
\end{eqnarray}
which explicitly displays a difference of two terms: the variance of the mean transverse momenta in events, and the
event-averaged variance of the transverse momentum in each event divided by its multiplicity. A technical simplification is that  
Eq.~(\ref{eq:my}) involves only single and not double sums in the event, which facilitates the computations. 

What is used in our comparisons to the data is the scaled measure $\langle \Delta p_T \Delta p_T \rangle^{1/2} / \langle \langle p_T \rangle \rangle$.

\section{Comparison to the ALICE data \label{sec:comp}}

\begin{figure}[tb]
\begin{center}
\includegraphics[width=0.38\textwidth]{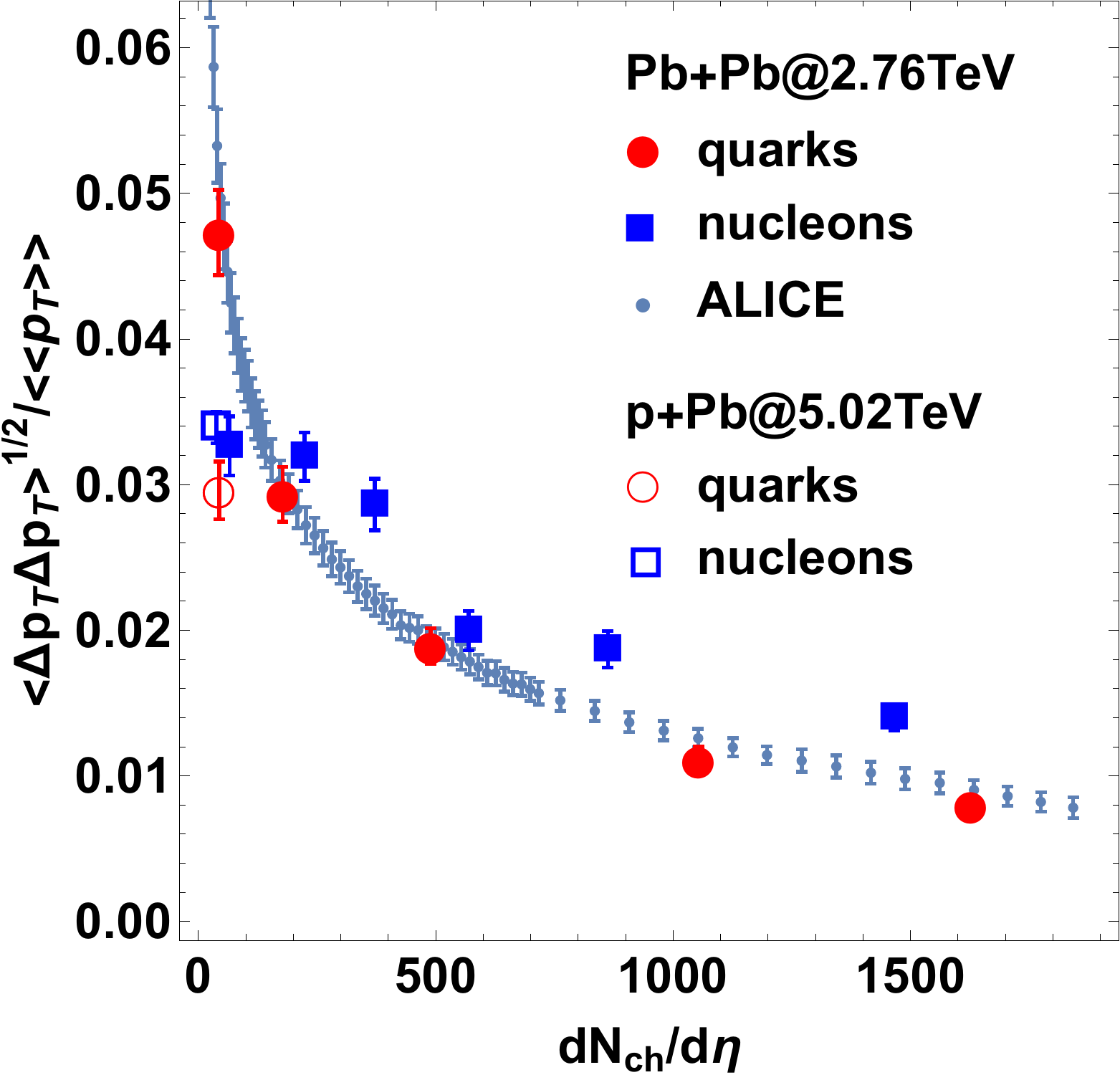} 
\end{center}
\vspace{-4mm}
\caption{The STAR measure of the transverse momentum fluctuations plotted vs charged hadron multiplicity. The simulations 
with the wounded quarks and nucleons are described in the text. The filled symbols correspond to the case of Pb+Pb collisions, whereas the 
empty symbols indicate our predictions for p+Pb collisions at centrality 0-3\%.
The experimental data for Pb+Pb case come from the ALICE Collaboration~\cite{Abelev:2014ckr}. \label{fig:ptfluct}}
\end{figure}

We start presenting our results with Fig.~\ref{fig:ptfluct}, where we compare $\langle \Delta p_T \Delta p_T \rangle^{1/2} / \langle \langle p_T \rangle \rangle$
obtained with simulations described in Sec.~\ref{sec:basic} to the experimental data. We use two models of the initial state: the wounded quark model of Eq.~(\ref{eq:wq})
and the mixed nucleon model of Eq.~(\ref{eq:wn}). The experimental data for Pb+Pb collisions at $\sqrt{s_{NN}}=2.76$~TeV come from the ALICE 
Collaboration~\cite{Abelev:2014ckr}. We note that the achieved agreement of the wounded quark model (circles) with the data is remarkable and extends over 
the whole multiplicity range, from central collisions (our right-most point corresponds to centrality 0-5\%)  to peripheral collisions of centrality 70-80\%.
This agreement is nontrivial, as a similar simulation but with the nucleon participants in the initial state is not nearly as good (squares): 
In this case, at highest multiplicities the data are overshot by about 50\%, whereas at low multiplicities the nucleon model predictions fall below the experiment.

\begin{figure}[tb]
\begin{center}
\includegraphics[width=0.38\textwidth]{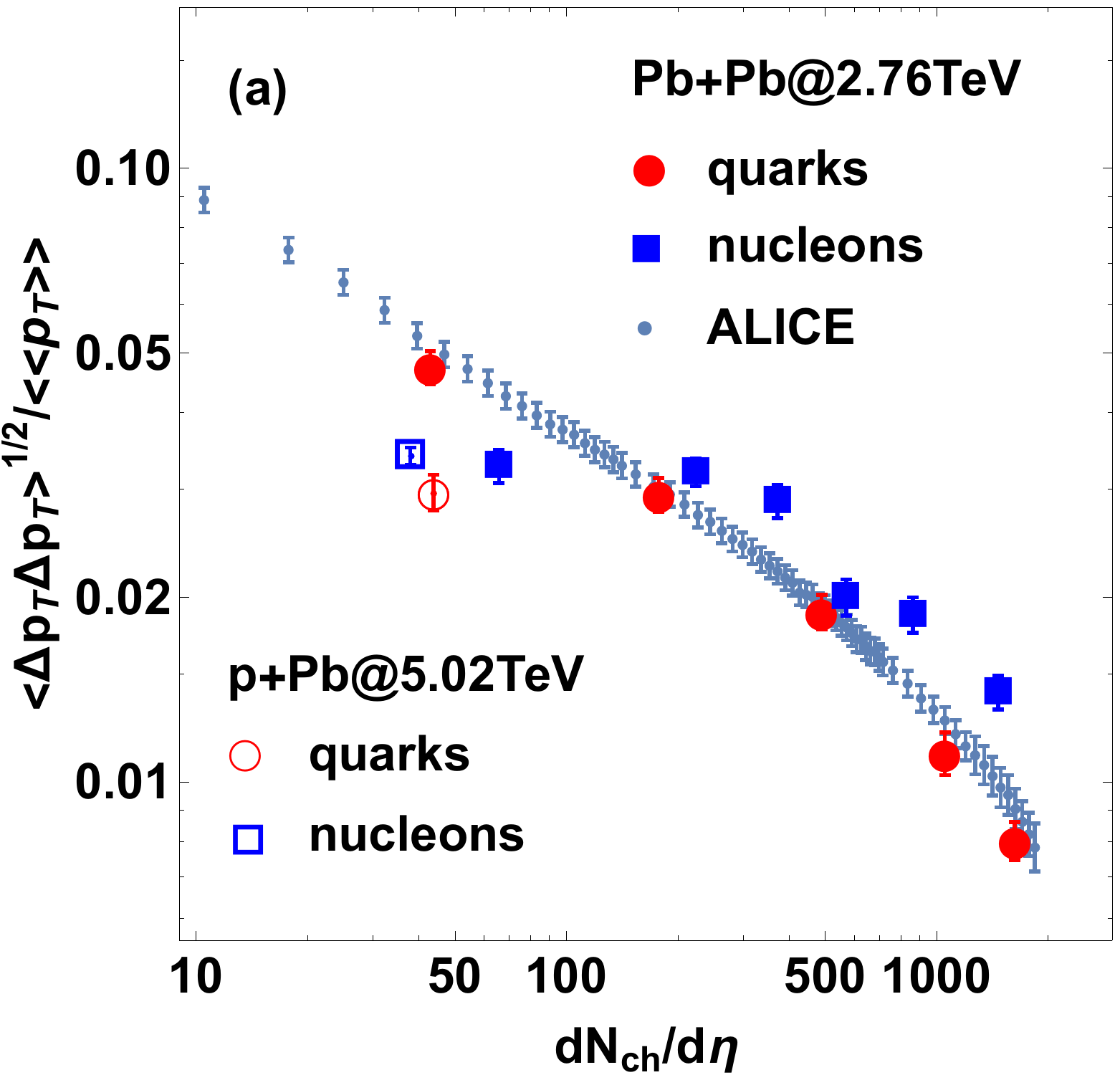} \\
\vspace{3mm}
\includegraphics[width=0.38\textwidth]{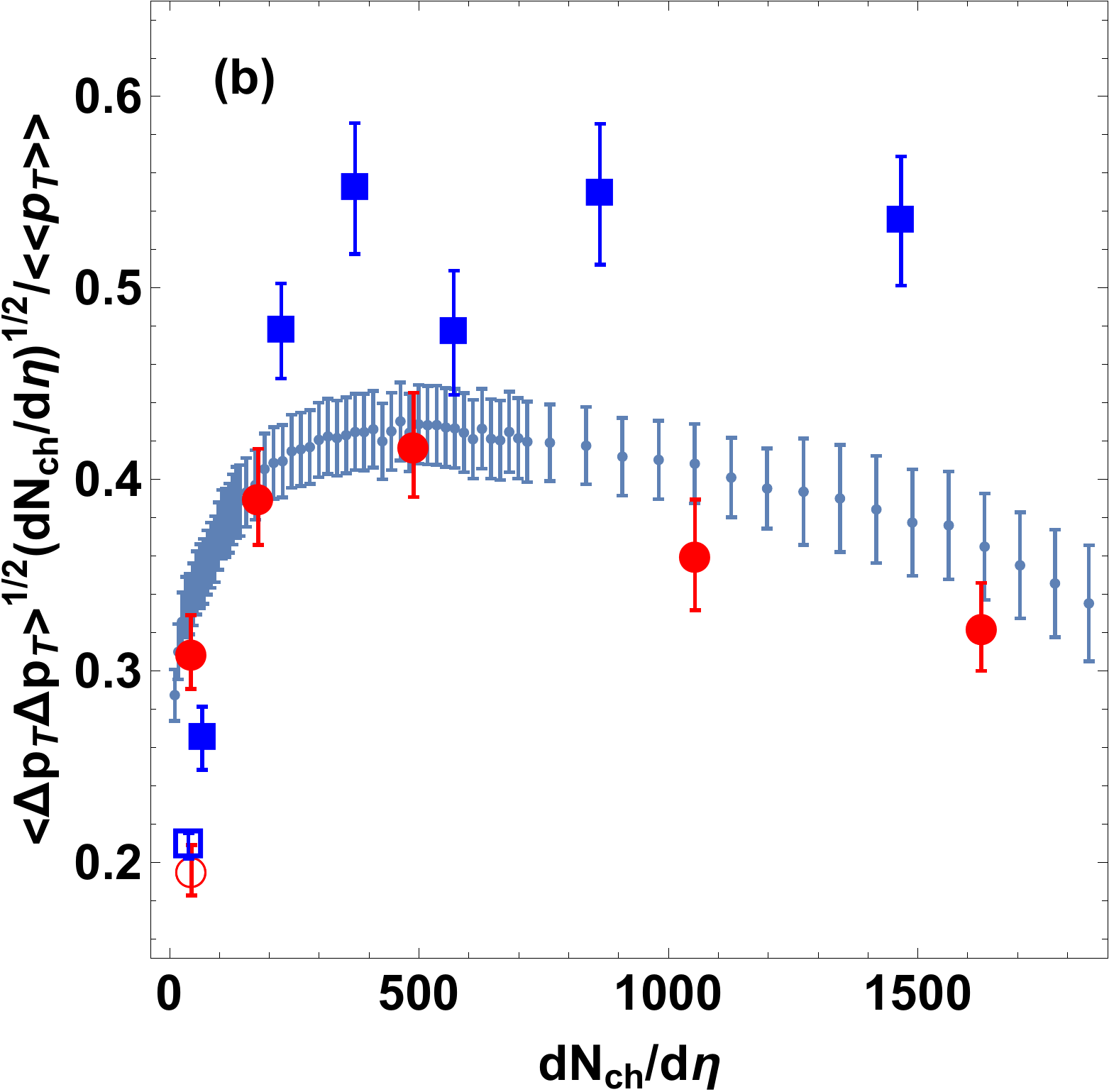} 
\end{center}
\vspace{-4mm}
\caption{The same as Fig.~\ref{fig:ptfluct} but (a)~plotted in the log-log scale and (b) for the 
ratio $\langle \Delta p_T \Delta p_T \rangle^{1/2} / [\langle \langle p_T \rangle \rangle (dN_{\rm ch}/d\eta)^{-1/2}]$
plotted in the log-linear scale. \label{fig:ptfluct_2}}
\end{figure}

We note that in the quark case the nontrivial experimental dependence on $dN_{\rm ch}/d\eta$ is reproduced.
As already pointed out in Ref.~\cite{Abelev:2014ckr}, this dependence does not follow a simple 
scaling with $(dN_{\rm ch}/d\eta)^{-1/2}$, which precludes the interpretation 
in terms of independent superposition of nucleon-nucleon collisions. 
For a better visualization of this issue, in Fig.~\ref{fig:ptfluct_2}(a) we show the result in the log-log scale. It can be 
clearly seen that the slope changes non-monotonically with the multiplicity, assuming largest negative values for most central collisions, then 
slightly flattening out, to increase a bit again for peripheral collisions. The same feature may be be read out from  Fig.~\ref{fig:ptfluct_2}(b), where 
following Ref.~\cite{Abelev:2014ckr} we examine the ratio  $\langle \Delta p_T \Delta p_T \rangle^{1/2} / [\langle \langle p_T \rangle \rangle (dN_{\rm ch}/d\eta)^{-1/2}]$, 
which is the relative slope with respect to the independent superposition case. 

In Figs.~\ref{fig:ptfluct} and \ref{fig:ptfluct_2} we also present our predictions for the p+Pb collisions at the LHC (empty symbols) for 
multiplicities corresponding to a high centrality 0-3\%. We recall that in such high-multiplicity collisions collectivity is expected to develop and
hydrodynamic description leads to appropriate phenomenology even in small systems (e.g. see Ref.~\cite{Bozek:2016jhf}).
These results may be confronted with future data analyses.

\section{Understanding the results}

This section is devoted to the understanding of the mechanism standing behind the agreement 
of the wounded quark+hydro approach exhibited in Figs.~\ref{fig:ptfluct} and \ref{fig:ptfluct_2}. 
The following discussion does not affect {\em per se} the results of the previous section, which were obtained in a robust 
way by just running the simulations, but it provides an interesting insight into the behavior of
hydrodynamics, in particular, its response to the initial conditions. 

\subsection{Initial size -- transverse momentum correlations  \label{sec:rpcorr}}

\begin{figure}[tb]
\includegraphics[width=0.38 \textwidth]{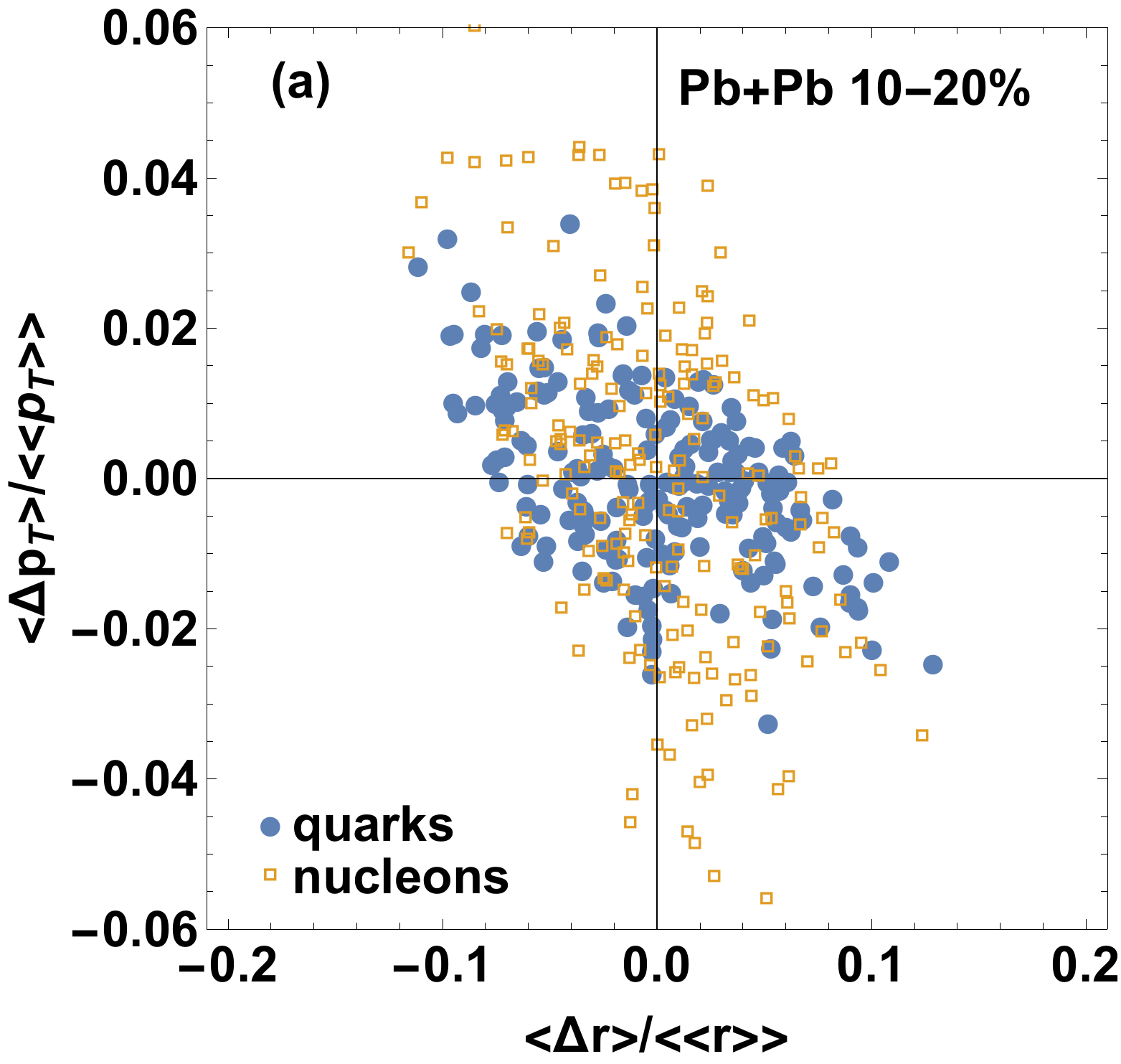} \\
 \includegraphics[width=0.38 \textwidth]{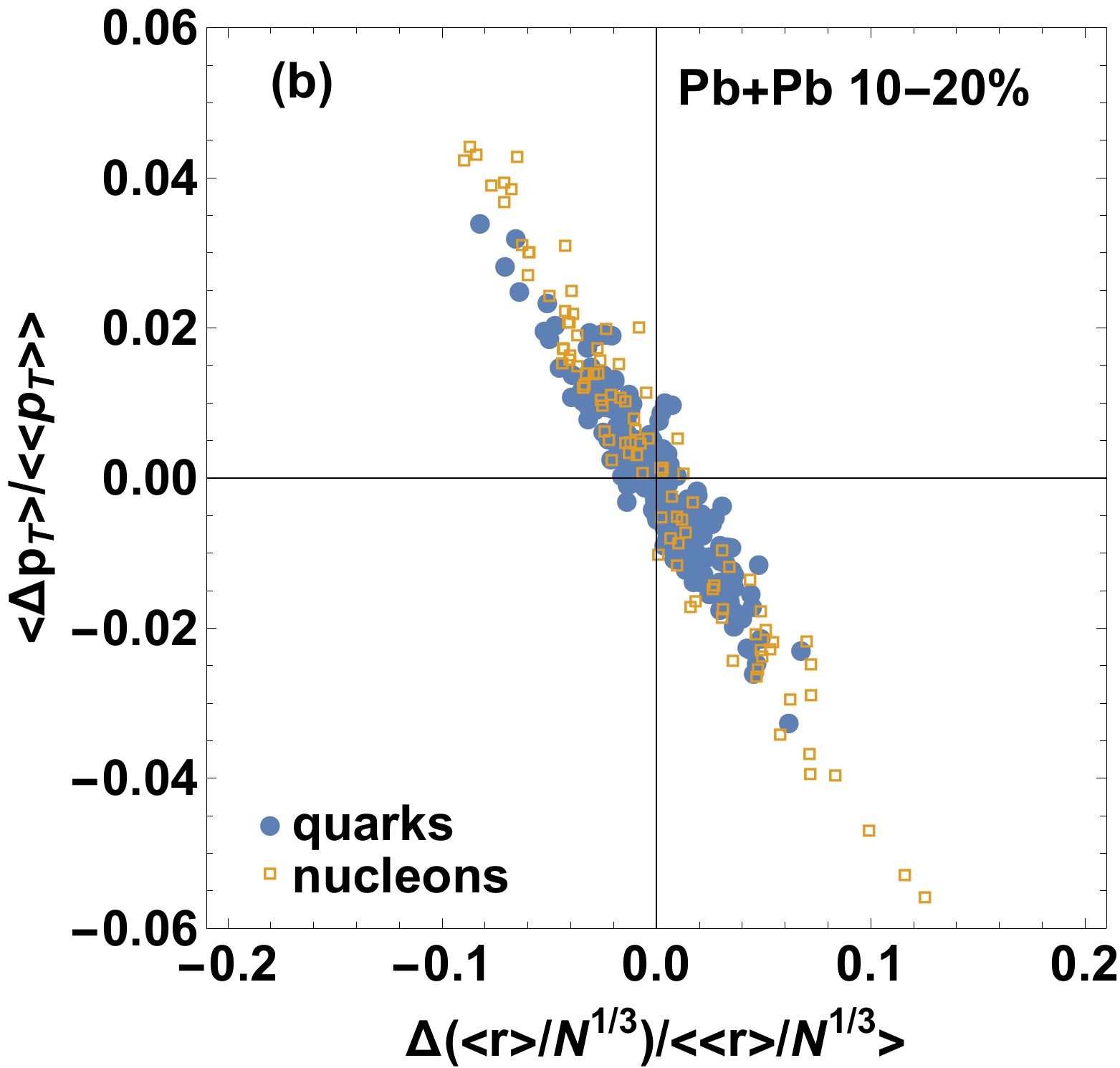} \\ 
\caption{
Scattered plot illustrating the correlation of the initial size and the transverse momentum in events for Pb+Pb collisions in  10-20\% centrality bin.
The results for the calculation using quark and nucleon Glauber models are denoted with  filled circles
and empty squares, respectively. Panel (a) shows the correlation between  the  r.m.s. radius and  
the  average transverse momentum in an event. Panel (b) displays the correlation between the rescaled r.m.s radius and the average transverse momentum. \label{fig:sc}}
\end{figure}

Generally, the obtained results depend on the fluctuations in the  initial condition and on their transmutation to 
the radial flow fluctuations by hydrodynamics. 
In the following we explore in detail how the size-flow transmutation is realized in our  hydrodynamic approach and 
also to what extent the size estimator of Eq.~\ref{eq:rd}
is universal in the sense that 
in a class of events with the same value of $\langle r \rangle$ (which otherwise may differ in shape or other radial moments) 
the resulting value of $\langle p_T \rangle$ is similar. To check it, we have generated scattered plots shown in Fig~\ref{fig:sc} (panel a), where we
plot $\langle \Delta r \rangle$ vs  $\langle \Delta p_T \rangle$, defined in event $k$ as
\begin{eqnarray}
\hspace{-4mm} 
\langle \Delta r \rangle_k = \langle r \rangle_k - \langle \langle r \rangle \rangle, \;\; \langle \Delta p_T \rangle_k = \langle p_T \rangle_k - \langle \langle p_T \rangle \rangle.
\end{eqnarray}
In this comparison the average transverse momentum for a given event 
is calculated by sampling the Cooper-Frye emission with a very large multiplicity 
($100$-$2000$ times the real event multiplicity). With such an oversampling, the obtained
values of the average transverse momentum in the event have a negligible 
component related to the finite multiplicity (the last term in Eq.~\ref{eq:my}), which 
simplifies the comparison.

\begin{figure}[tb]
\begin{center}
\includegraphics[width=0.38 \textwidth]{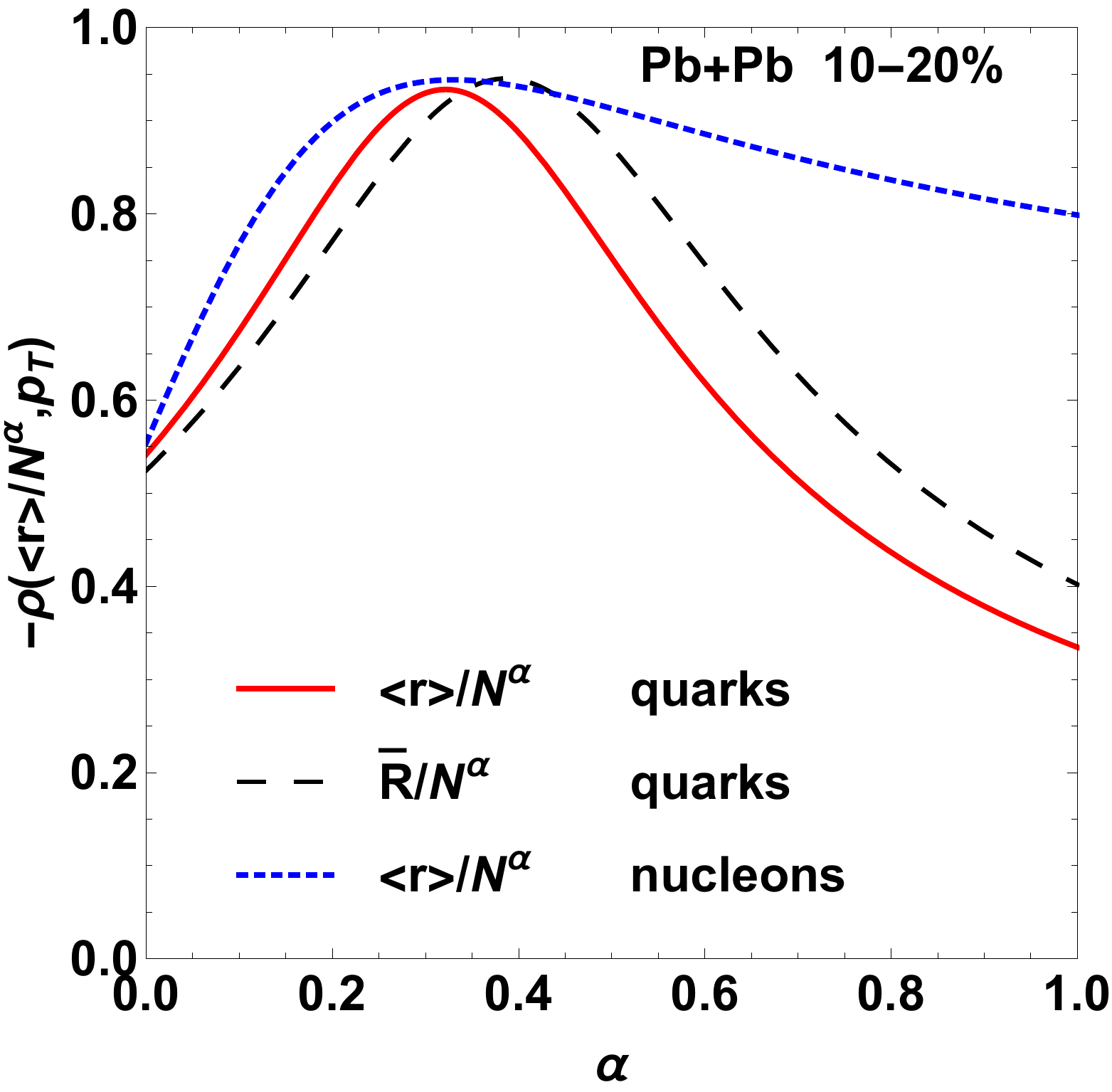}
\end{center}
\caption{Dependence of the anticorrelation between the  rescaled fireball size $<r>/N^\alpha$  and the  average $\langle p_T \rangle$ in the event 
on the scaling power $\alpha$ for 10-20\% centrality Pb+Pb collisions. \label{fig:cora}}
\end{figure}

Before further discussion, let us mention 
a rather technical issue which will, however, lead to interesting conclusions. 
Our samples correspond to centrality classes with the standard widths of 5\% for the most central and 10\% for 
more peripheral collisions. Within such rather wide bins, necessary for sufficient statistics, the multiplicity of events  
fluctuates. One trivially expects that on the average larger size fireballs will also ultimately produce more hadrons. 
To remove this trivial effect of centrality fluctuations one should take narrow centrality classes, as was possible in the experiment~\cite{Abelev:2014ckr}.
Our limited statistics does not allow for this remedy, however, one can resort to another method. We find that in the 
applied Glauber model~\cite{Broniowski:2007nz,*Rybczynski:2013yba} the increase of the average fireball size with the number of participants 
scales approximately as $\langle r \rangle \propto N^{0.2-0.5}$.
It means that within a given centrality class a large part of the size variation comes from a change in the initial entropy. 
We improve the size
measure of Eq.~(\ref{eq:rd}) by scaling it with a power of the entropy in the event (or a related measure such as the number of wounded participants), 
\begin{eqnarray}
\langle r \rangle_k \to  \langle r \rangle_k /N_k^\alpha, \label{eq:rN}
\end{eqnarray}
where $\alpha$ is a numerically adjusted parameter. In Fig. \ref{fig:cora} we display the 
correlation coefficient between $\langle \Delta r\rangle / N^\alpha$ and $\langle \Delta p_T \rangle$ for events in the $10-20$\% centrality class as a function of  $\alpha$.
Both for the models with quark and nucleon participants a maximum correlation occurs for $\alpha\simeq 1/3$ (dashed and dashed-dotted lines in Fig. \ref{fig:cora}). We 
have also tested the correlation in the case when the size  is estimated using the asymmetric 
variable $\bar{R}$ of Eq.~\ref{eq:barr}. In all the cases the correlation is strongest around $\alpha=0.3-0.4$, with similar values of the correlation coefficient.
 For the calculation with the quark Glauber model the value of $\alpha$ where the correlation is maximal 
 changes weakly with centrality, whereas for the nucleon model it increases up to $\alpha\simeq 0.42$ 
in peripheral collisions.

The estimator of (\ref{eq:rN}) 
works much better than the standard definition of Eq.~\ref{eq:rd}, as can be promptly seen from panel b)
 in Fig.~\ref{fig:sc}. 
The improvement occurs for both the wounded quark model (filled symbols) and the nucleon case (empty symbols).
We have checked the the rescaled estimator yields a much stronger correlation with the final transverse momentum
 for all centralities in Pb+Pb and for central  p+Pb collisions. 

The rescaled radius (\ref{eq:rN}) is very similar to the estimator of the strength of the subleading component
in the principal component analysis of the transverse momentum spectra by Mazeliauskas and Teaney~\cite{Mazeliauskas:2015efa},
who use as an estimator the combination 
\begin{equation}
\langle \Delta p_T\rangle \propto \beta \epsilon_{0,0} +  \epsilon_{0,2} \ ,
\label{eq:MT}
\end{equation}
where 
\begin{equation}
\epsilon_{m,n} e^{i m \Psi_{m,n}} = 
- \frac{\int dx\,dy\, s_k(x,y) (x^2+y^2)^{n/2} e^{im\phi}}{\langle \int dx\,dy\, s_k(x,y)\rangle \langle \langle r \rangle \rangle^{n/2}} 
\label{eq:epsil}
\end{equation}
For the calculation in the wounded quark model 
we find that the largest correlation with the transverse momentum 
with the estimator (\ref{eq:MT}) occurs 
for $\beta \simeq -5/3$. To linear order in deviations from average values in the centrality bin,
this is equivalent to the rescaled r.m.s. radius (\ref{eq:rN}) with $\alpha\simeq 1/3$.

\begin{figure}[tb]
\begin{center}
\includegraphics[width=0.38\textwidth]{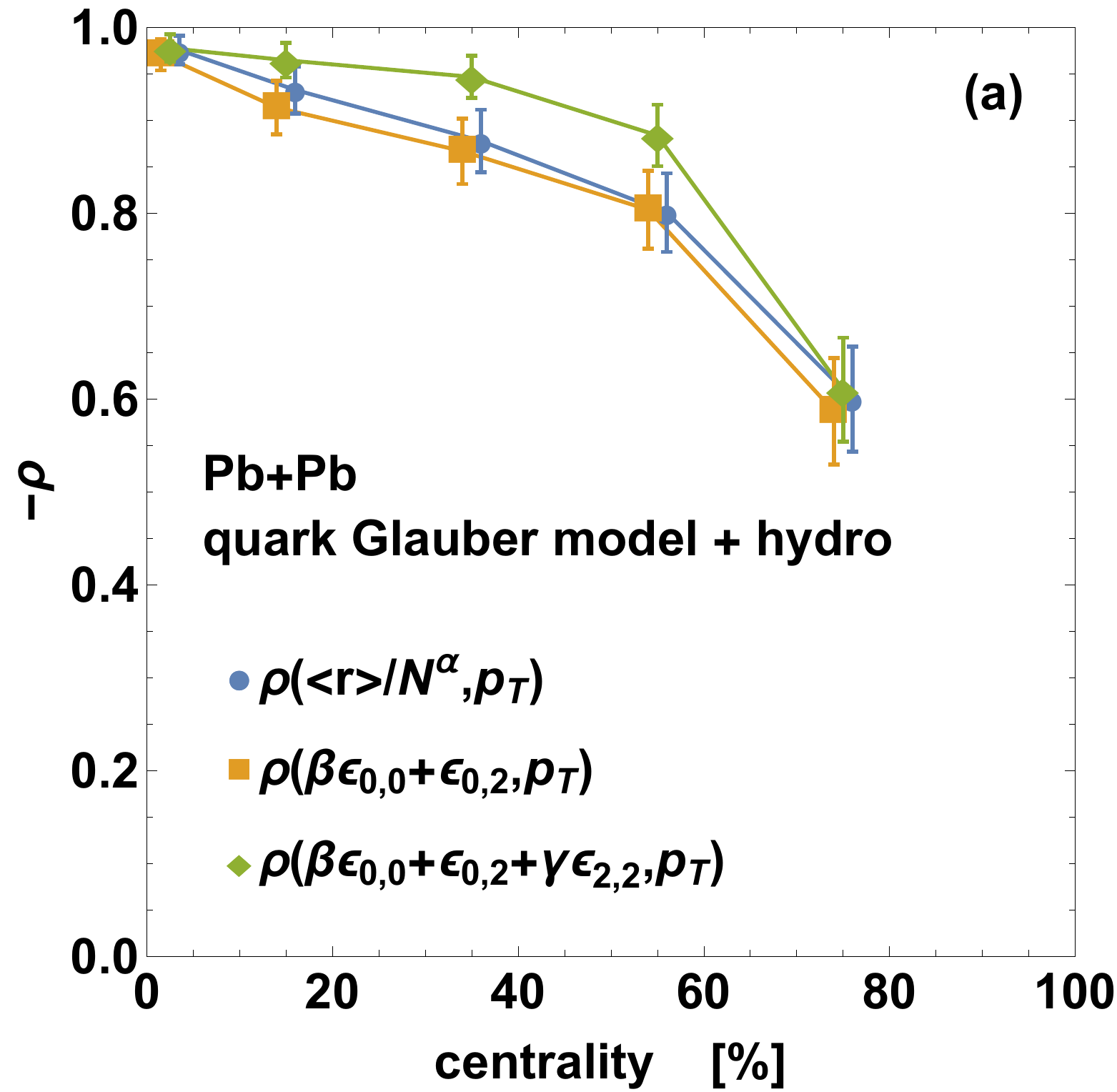}\\
\vspace{2mm}
\includegraphics[width=0.38 \textwidth]{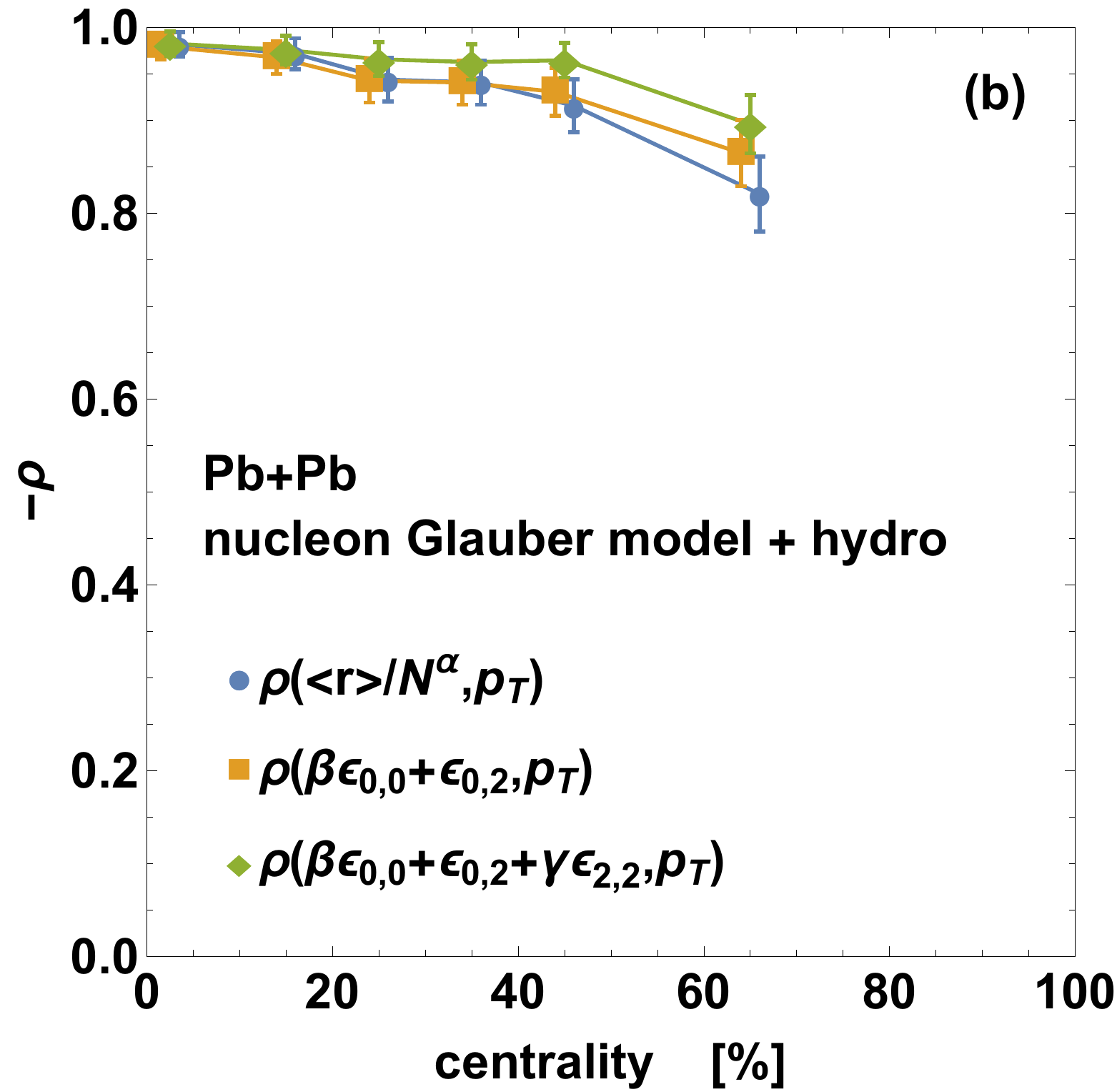}
\end{center}
\caption{Event-by-event anticorrelation coefficient of 
the estimators of size described in the text and the transverse momentum $\langle p_T \rangle$, plotted 
versus the centrality of the collision. \label{fig:corr}}
\end{figure}

The anticorrelation coefficient for the estimator $\langle r \rangle/N^\alpha$ and for the linear combination $\beta \epsilon_{0,0}+\epsilon_{0,2}$ 
is displayed in Fig.~\ref{fig:corr} for the initial state given by
the quark (panel a) and nucleon (panel b) models.
For both models we notice that the estimator $\langle r \rangle/N^\alpha$ and the estimator~(\ref{eq:MT}) give very similar results, as expected.
Quantitatively, the two models have different correlations. In the nucleon model the anticorrelation is very strong for all
centralities. In the quark model 
for the most central collisions the anticorrelation is essentially perfect, with the correlation coefficient very close to $-1$. It gradually increases to 
about -0.6 at \mbox{$c=80\%$}.
Estimators based on $\bar{R}$ yield similar results. A significant improvement in the quality of the 
estimator for the quark model is visible only when using a formula including initial eccentricity (diamonds in Fig.~\ref{fig:corr}(a))
\begin{equation}
\langle \Delta p_T\rangle \propto \beta \epsilon_{0,0} +  \epsilon_{0,2} +\gamma  \epsilon_{2,2}\ .
\label{eq:MT2}
\end{equation}
With the two-parameter estimator (\ref{eq:MT2}), the anticorrelation with the final
momentum is strong except for the most peripheral bin. 
The use of this  more general formula gives no significant improvement in the prediction of the final $\langle p_T \rangle$ 
for the nucleon model (diamonds in Fig.~\ref{fig:corr}(b)).

\subsection{Size fluctuations in the initial state}

\begin{figure}[tb]
\begin{center}
\includegraphics[width=0.38\textwidth]{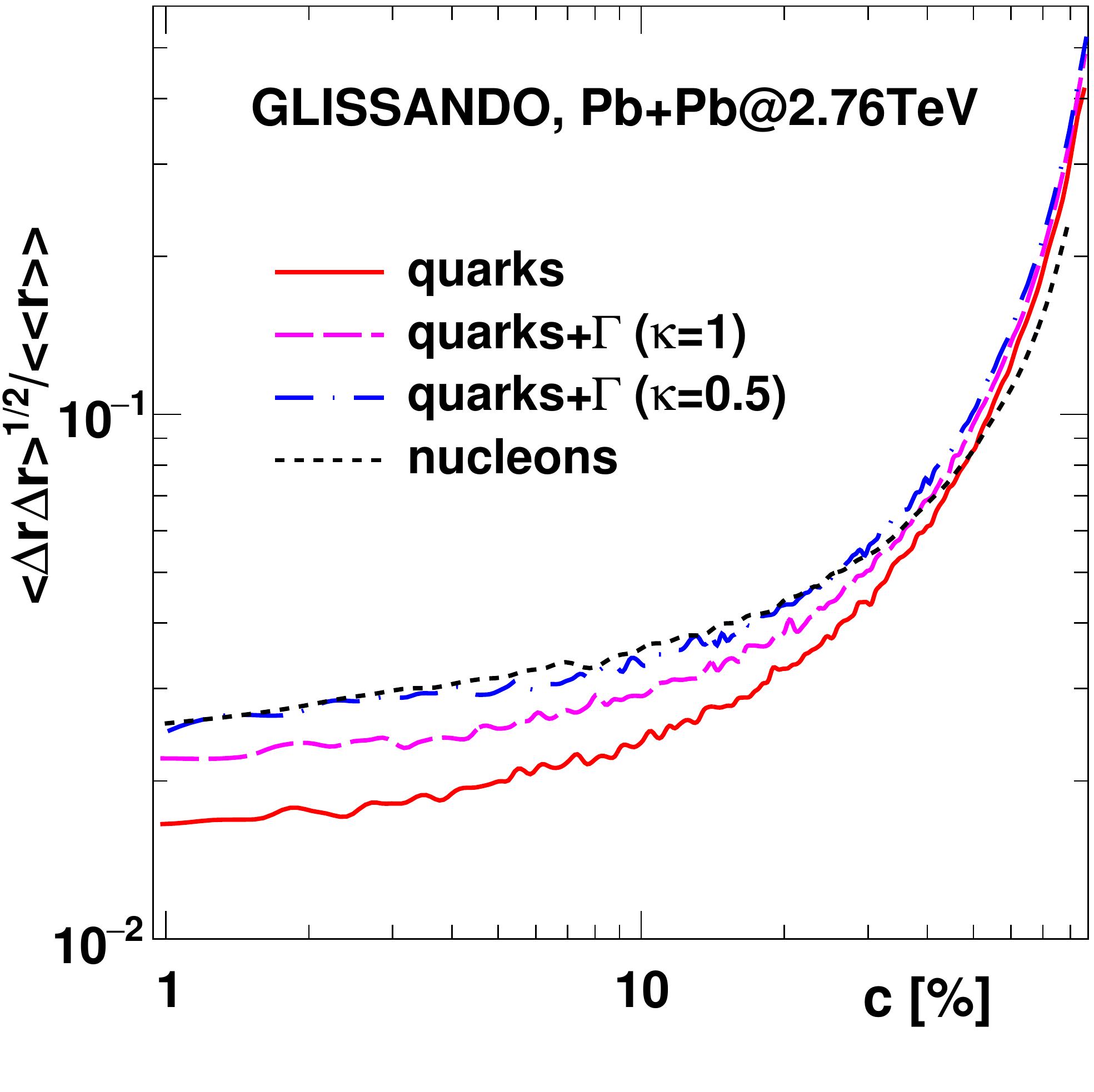}
\end{center}
\vspace{-10mm}
\caption{Size fluctuations plotted as functions of centrality in various models of the initial state. See text for details.  \label{fig:gliss}}
\end{figure}

We now present a closer look at the systematics of the initial condition generated in our approach with
{\tt GLISSANDO}~\cite{Broniowski:2007nz,*Rybczynski:2013yba}. We also explore a potentially relevant 
ingredient, namely, the possible additional fluctuations in the initial state. The idea here is based on an intuitive expectation that 
an elementary collision need not always deposit the same amount of entropy, but the quantity may fluctuate. In fact, in small 
systems such fluctuations are needed to describe the multiplicity fluctuations of the produced hadrons.
In Ref.~\cite{Broniowski:2007nz} we have proposed that the strength of the Glauber sources should  fluctuate according to 
the $\Gamma$ distribution, which then upon folding with the Poissonian distribution due to hadronization and detector acceptance, 
yields the negative binomial distribution, efficient in fitting the data. It is not a priori clear how much of these extra fluctuations should be present in the 
initial state in the A+A collisions. The parameter which controls their magnitude is $\kappa$, which equals to the ratio of the mean squared to the variance of the 
$\Gamma$ distribution of the source strengths . 

The results are shown in Fig.~\ref{fig:gliss}, where we plot the scaled standard deviation  of the transverse size  given by Eq.~(\ref{eq:rd}) as a function of centrality. Very narrow centrality bins 
are determined independently for each model, such that the models can be compared in a uniform way. We note several facts. First, size fluctuations 
from the wounded quark model without fluctuations are, at low centralities, significantly below the nucleon model. 
This explains the fact that 
the quark simulations presented in Sec.~\ref{sec:comp} work well, as opposed to the nucleon case. Second, increasing the 
overlaid fluctuations (by 
decreasing $\kappa$) lead to increased size fluctuations, as intuitively expected.

\subsection{Insensitivity to variants of hadronization}

\begin{figure}[tb]
\begin{center}
\includegraphics[width=0.38\textwidth]{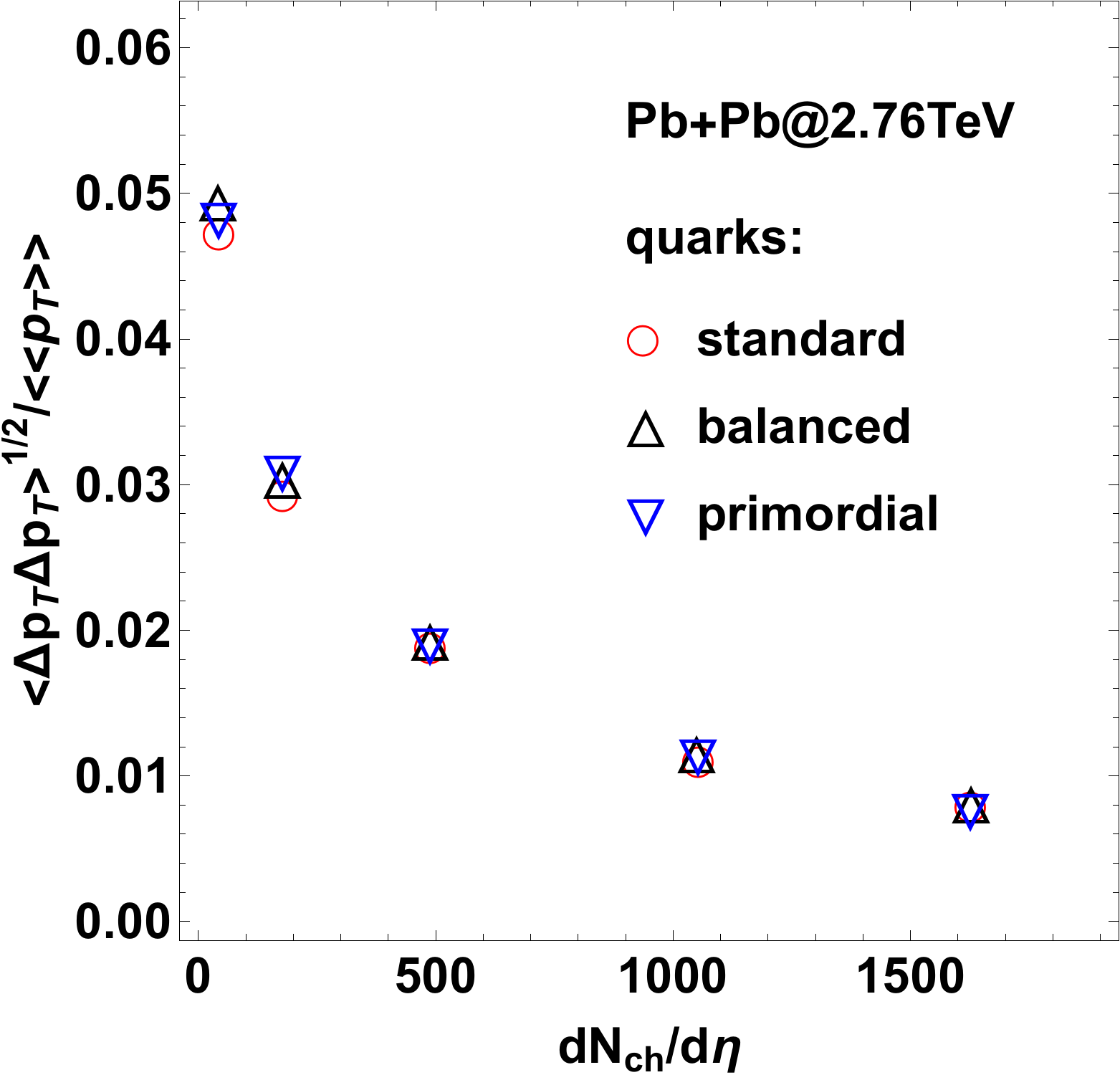}
\end{center}
\vspace{-5mm}
\caption{Same as in Fig.~\ref{fig:ptfluct} for the wounded quark case with various models of hadronization.  \label{fig:bal}}
\end{figure}

In a final check we show a remarkable insensitivity of our simulations to models of 
hadronization. In Fig.~\ref{fig:bal} we show the results from a standard
{\tt THERMINATOR} hadronization model taking into account charged pions, kaons, protons and antiprotons after decays of all resonances (circles), 
from a model which incorporates late charge 
balancing as implemented in Ref.~\cite{Bozek:2012en} (up triangles), and from 
the primordial (before resonance decays) charged pions, kaons, protons and antiprotons (down triangles).
The near-equality of all variants reflects the robustness of the STAR correlation measure applied in our analysis. 

\section{Conclusions}

Fluctuations of the average transverse momentum of particles emitted in relativistic heavy-ion collisions 
are calculated using a viscous hydrodynamic model. Event-by-event fluctuations of the fireball lead to
 event-by-event fluctuations of the average transverse momentum. The initial state obtained in the 
 Glauber Monte Carlo model with quark degrees of freedom describes properly the $\langle p_T \rangle$ fluctuations 
in Pb+Pb collisions measured by the ALICE Collaboration. The data are well described in a wide range of 
centralities, from $0-5$\% to $70-80$\%. On the other hand, the calculation using the initial 
state with nucleon degrees of freedom overestimates the observed data at central collisions.
The dependence of the experimentally observed fluctuations on centrality departs from the simple scaling  
$(dN_{\rm ch}/d\eta)^{-1/2}$, holding for the independent superposition model. 
This nontrivial centrality dependence is well described by our calculation. 
Predicted momentum fluctuations in p+Pb collisions are smaller than in Pb+Pb collisions at
the same multiplicity of produced charged hadrons.

The average transverse flow in an event is correlated with the initial r.m.s radius 
scaled by a power of the initial entropy.  We confirm the results of Ref.~\cite{Mazeliauskas:2015efa} that this  
correlations is almost perfect in the model with nucleon degrees of freedom. In the calculation using the 
wounded quark  model the correlation is somewhat weaker. 
An improvement of the correlations of the estimators with the final transverse
momentum is possible adding a term to the estimator formula with the initial eccentricity.

The comparison of model results using primordial particles, particles including resonance decay products, 
or the calculation with imposed late-stage local charge conservation effects show that the resulting effects 
are very small in transverse momentum fluctuations. 
However, other sources of fluctuations can increase the observed $\langle p_T \rangle$ fluctuations. 
We show that additional fluctuations in the entropy deposition in the initial state may 
give rise to a significant increase of $\langle p_T \rangle$ fluctuations. 
The measurements of the ALICE Collaboration can thus provide an upper bound on the the strength of 
such entropy fluctuations in the initial state.

\begin{acknowledgments}

Research supported by the Polish Ministry of Science and Higher Education (MNiSW), by the National
Science Centre grant 2015/17/B/ST2/00101 (PB) and grant 2015/19/B/ST2/00937 
(WB), as well as by PL-Grid Infrastructure. 

\end{acknowledgments}

\bibliography{../../hydr}

\begin{thebibliography}{10}%
\makeatletter
\providecommand \@ifxundefined [1]{%
 \ifx #1\undefined \expandafter \@firstoftwo
 \else \expandafter \@secondoftwo
\fi
}%
\providecommand \@ifnum [1]{%
 \ifnum #1\expandafter \@firstoftwo
 \else \expandafter \@secondoftwo
\fi
}%
\providecommand \enquote [1]{``#1''}%
\providecommand \bibnamefont  [1]{#1}%
\providecommand \bibfnamefont [1]{#1}%
\providecommand \citenamefont [1]{#1}%
\providecommand\href[0]{\@sanitize\@href}%
\providecommand\@href[1]{\endgroup\@@startlink{#1}\endgroup\@@href}%
\providecommand\@@href[1]{#1\@@endlink}%
\providecommand \@sanitize [0]{\begingroup\catcode`\&12\catcode`\#12\relax}%
\@ifxundefined \pdfoutput {\@firstoftwo}{%
 \@ifnum{\z@=\pdfoutput}{\@firstoftwo}{\@secondoftwo}%
}{%
 \providecommand\@@startlink[1]{\leavevmode\special{html:<a href="#1">}}%
 \providecommand\@@endlink[0]{\special{html:</a>}}%
}{%
 \providecommand\@@startlink[1]{%
  \leavevmode
  \pdfstartlink
   attr{/Border[0 0 1 ]/H/I/C[0 1 1]}%
   user{/Subtype/Link/A<</Type/Action/S/URI/URI(#1)>>}%
  \relax
 }%
 \providecommand\@@endlink[0]{\pdfendlink}%
}%
\providecommand \url  [0]{\begingroup\@sanitize \@url }%
\providecommand \@url [1]{\endgroup\@href {#1}{\urlprefix}}%
\providecommand \urlprefix [0]{URL }%
\providecommand \Eprint[0]{\href }%
\@ifxundefined \urlstyle {%
  \providecommand \doi [1]{doi:\discretionary{}{}{}#1}%
}{%
  \providecommand \doi [0]{doi:\discretionary{}{}{}\begingroup
  \urlstyle{rm}\Url }%
}%
\providecommand \doibase [0]{http://dx.doi.org/}%
\providecommand \Doi[1]{\href{\doibase#1}}%
\providecommand \bibAnnote [3]{%
  \BibitemShut{#1}%
  \begin{quotation}\noindent
    \textsc{Key:}\ #2\\\textsc{Annotation:}\ #3%
  \end{quotation}%
}%
\providecommand \bibAnnoteFile [2]{%
  \IfFileExists{#2}{\bibAnnote {#1} {#2} {\input{#2}}}{}%
}%
\providecommand \typeout [0]{\immediate \write \m@ne }%
\providecommand \selectlanguage [0]{\@gobble}%
\providecommand \bibinfo [0]{\@secondoftwo}%
\providecommand \bibfield [0]{\@secondoftwo}%
\providecommand \translation [1]{[#1]}%
\providecommand \BibitemOpen[0]{}%
\providecommand \bibitemStop [0]{}%
\providecommand \bibitemNoStop [0]{.\EOS\space}%
\providecommand \EOS [0]{\spacefactor3000\relax}%
\providecommand \BibitemShut [1]{\csname bibitem#1\endcsname}%
\bibitem{Broniowski:2009fm}%
  \BibitemOpen
  \bibfield{author}{%
  \bibinfo {author} {\bibfnamefont{W.}~\bibnamefont{Broniowski}}, \bibinfo
  {author} {\bibfnamefont{M.}~\bibnamefont{Chojnacki}},\ and\ \bibinfo {author}
  {\bibfnamefont{L.}~\bibnamefont{Obara}},\ }%
  \bibfield{journal}{%
  \Doi{10.1103/PhysRevC.80.051902}{\bibinfo {journal} {Phys. Rev.}}\ }%
  \textbf{\bibinfo {volume} {C80}},\ \bibinfo {pages} {051902} (\bibinfo {year}
  {2009})%
  \bibAnnoteFile{NoStop}{Broniowski:2009fm}%
\bibitem{Bozek:2012fw}%
  \BibitemOpen
  \bibfield{author}{%
  \bibinfo {author} {\bibfnamefont{P.}~\bibnamefont{Bo\.zek}}\ and\ \bibinfo
  {author} {\bibfnamefont{W.}~\bibnamefont{Broniowski}},\ }%
  \bibfield{journal}{%
  \bibinfo {journal} {Phys. Rev.}\ }%
  \textbf{\bibinfo {volume} {C85}},\ \bibinfo {pages} {044910} (\bibinfo {year}
  {2012})%
  \bibAnnoteFile{NoStop}{Bozek:2012fw}%
\bibitem{Bialas:1977en}%
  \BibitemOpen
  \bibfield{author}{%
  \bibinfo {author} {\bibfnamefont{A.}~\bibnamefont{Bia\l{}as}}, \bibinfo
  {author} {\bibfnamefont{W.}~\bibnamefont{Czy\.z}},\ and\ \bibinfo {author}
  {\bibfnamefont{W.}~\bibnamefont{Furma\'nski}},\ }%
  \bibfield{journal}{%
  \bibinfo {journal} {Acta Phys. Polon.}\ }%
  \textbf{\bibinfo {volume} {B8}},\ \bibinfo {pages} {585} (\bibinfo {year}
  {1977})%
  \bibAnnoteFile{NoStop}{Bialas:1977en}%
\bibitem{Bialas:1977xp}%
  \BibitemOpen
  \bibfield{author}{%
  \bibinfo {author} {\bibfnamefont{A.}~\bibnamefont{Bia\l{}as}}, \bibinfo
  {author} {\bibfnamefont{K.}~\bibnamefont{Fia\l{}kowski}}, \bibinfo {author}
  {\bibfnamefont{W.}~\bibnamefont{S\l{}omi\'nski}},\ and\ \bibinfo {author}
  {\bibfnamefont{M.}~\bibnamefont{Zieli\'nski}},\ }%
  \bibfield{journal}{%
  \bibinfo {journal} {Acta Phys. Polon.}\ }%
  \textbf{\bibinfo {volume} {B8}},\ \bibinfo {pages} {855} (\bibinfo {year}
  {1977})%
  \bibAnnoteFile{NoStop}{Bialas:1977xp}%
\bibitem{Bialas:1978ze}%
  \BibitemOpen
  \bibfield{author}{%
  \bibinfo {author} {\bibfnamefont{A.}~\bibnamefont{Bia\l{}as}}\ and\ \bibinfo
  {author} {\bibfnamefont{W.}~\bibnamefont{Czy\.z}},\ }%
  \bibfield{journal}{%
  \bibinfo {journal} {Acta Phys. Polon.}\ }%
  \textbf{\bibinfo {volume} {B10}},\ \bibinfo {pages} {831} (\bibinfo {year}
  {1979})%
  \bibAnnoteFile{NoStop}{Bialas:1978ze}%
\bibitem{Anisovich:1977av}%
  \BibitemOpen
  \bibfield{author}{%
  \bibinfo {author} {\bibfnamefont{V.~V.}\ \bibnamefont{Anisovich}}, \bibinfo
  {author} {\bibfnamefont{{\relax Yu}.~M.}\ \bibnamefont{Shabelski}},\ and\
  \bibinfo {author} {\bibfnamefont{V.~M.}\ \bibnamefont{Shekhter}},\ }%
  \bibfield{journal}{%
  \Doi{10.1016/0550-3213(78)90237-7}{\bibinfo {journal} {Nucl. Phys.}}\ }%
  \textbf{\bibinfo {volume} {B133}},\ \bibinfo {pages} {477} (\bibinfo {year}
  {1978})%
  \bibAnnoteFile{NoStop}{Anisovich:1977av}%
\bibitem{Abelev:2014ckr}%
  \BibitemOpen
  \bibfield{author}{%
  \bibinfo {author} {\bibfnamefont{B.~B.}\ \bibnamefont{Abelev}} \emph{et~al.}
  (\bibinfo {collaboration} {ALICE}),\ }%
  \bibfield{journal}{%
  \Doi{10.1140/epjc/s10052-014-3077-y}{\bibinfo {journal} {Eur. Phys. J.}}\ }%
  \textbf{\bibinfo {volume} {C74}},\ \bibinfo {pages} {3077} (\bibinfo {year}
  {2014})%
  \bibAnnoteFile{NoStop}{Abelev:2014ckr}%
\bibitem{Lacey:2016hqy}%
  \BibitemOpen
  \bibfield{author}{%
  \bibinfo {author} {\bibfnamefont{R.~A.}\ \bibnamefont{Lacey}}, \bibinfo
  {author} {\bibfnamefont{P.}~\bibnamefont{Liu}}, \bibinfo {author}
  {\bibfnamefont{N.}~\bibnamefont{Magdy}}, \bibinfo {author}
  {\bibfnamefont{M.}~\bibnamefont{Csanád}}, \bibinfo {author}
  {\bibfnamefont{B.}~\bibnamefont{Schweid}}, \bibinfo {author}
  {\bibfnamefont{N.~N.}\ \bibnamefont{Ajitanand}}, \bibinfo {author}
  {\bibfnamefont{J.}~\bibnamefont{Alexander}},\ and\ \bibinfo {author}
  {\bibfnamefont{R.}~\bibnamefont{Pak}}}%
   (\bibinfo {year} {2016}),\
  \Eprint{http://arxiv.org/abs/1601.06001}{arXiv:1601.06001 [nucl-ex]}%
  \bibAnnoteFile{NoStop}{Lacey:2016hqy}%
\bibitem{Bozek:2016kpf}%
  \BibitemOpen
  \bibfield{author}{%
  \bibinfo {author} {\bibfnamefont{P.}~\bibnamefont{Bo{\.z}ek}}, \bibinfo
  {author} {\bibfnamefont{W.}~\bibnamefont{Broniowski}},\ and\ \bibinfo
  {author} {\bibfnamefont{M.}~\bibnamefont{Rybczy{\'n}ski}},\ }%
  \bibfield{journal}{%
  \Doi{10.1103/PhysRevC.94.014902}{\bibinfo {journal} {Phys. Rev.}}\ }%
  \textbf{\bibinfo {volume} {C94}},\ \bibinfo {pages} {014902} (\bibinfo {year}
  {2016})%
  \bibAnnoteFile{NoStop}{Bozek:2016kpf}%
\bibitem{Gazdzicki:1992ri}%
  \BibitemOpen
  \bibfield{author}{%
  \bibinfo {author} {\bibfnamefont{M.}~\bibnamefont{Gazdzicki}}\ and\ \bibinfo
  {author} {\bibfnamefont{S.}~\bibnamefont{Mrowczynski}},\ }%
  \bibfield{journal}{%
  \bibinfo {journal} {Z. Phys.}\ }%
  \textbf{\bibinfo {volume} {C54}},\ \bibinfo {pages} {127} (\bibinfo {year}
  {1992})%
  \bibAnnoteFile{NoStop}{Gazdzicki:1992ri}%
\bibitem{Stodolsky:1995ds}%
  \BibitemOpen
  \bibfield{author}{%
  \bibinfo {author} {\bibfnamefont{L.}~\bibnamefont{Stodolsky}},\ }%
  \bibfield{journal}{%
  \bibinfo {journal} {Phys. Rev. Lett.}\ }%
  \textbf{\bibinfo {volume} {75}},\ \bibinfo {pages} {1044} (\bibinfo {year}
  {1995})%
  \bibAnnoteFile{NoStop}{Stodolsky:1995ds}%
\bibitem{Shuryak:1997yj}%
  \BibitemOpen
  \bibfield{author}{%
  \bibinfo {author} {\bibfnamefont{E.~V.}\ \bibnamefont{Shuryak}},\ }%
  \bibfield{journal}{%
  \bibinfo {journal} {Phys. Lett.}\ }%
  \textbf{\bibinfo {volume} {B423}},\ \bibinfo {pages} {9} (\bibinfo {year}
  {1998})%
  \bibAnnoteFile{NoStop}{Shuryak:1997yj}%
\bibitem{Mrowczynski:1997kz}%
  \BibitemOpen
  \bibfield{author}{%
  \bibinfo {author} {\bibfnamefont{S.}~\bibnamefont{Mrowczynski}},\ }%
  \bibfield{journal}{%
  \bibinfo {journal} {Phys. Lett.}\ }%
  \textbf{\bibinfo {volume} {B430}},\ \bibinfo {pages} {9} (\bibinfo {year}
  {1998})%
  \bibAnnoteFile{NoStop}{Mrowczynski:1997kz}%
\bibitem{Liu:1998xf}%
  \BibitemOpen
  \bibfield{author}{%
  \bibinfo {author} {\bibfnamefont{F.}~\bibnamefont{Liu}}, \bibinfo {author}
  {\bibfnamefont{A.}~\bibnamefont{Tai}}, \bibinfo {author}
  {\bibfnamefont{M.}~\bibnamefont{Gazdzicki}},\ and\ \bibinfo {author}
  {\bibfnamefont{R.}~\bibnamefont{Stock}},\ }%
  \bibfield{journal}{%
  \Doi{10.1007/s100529900002}{\bibinfo {journal} {Eur.Phys.J.}}\ }%
  \textbf{\bibinfo {volume} {C8}},\ \bibinfo {pages} {649} (\bibinfo {year}
  {1999})%
  \bibAnnoteFile{NoStop}{Liu:1998xf}%
\bibitem{Voloshin:1999yf}%
  \BibitemOpen
  \bibfield{author}{%
  \bibinfo {author} {\bibfnamefont{S.~A.}\ \bibnamefont{Voloshin}}, \bibinfo
  {author} {\bibfnamefont{V.}~\bibnamefont{Koch}},\ and\ \bibinfo {author}
  {\bibfnamefont{H.~G.}\ \bibnamefont{Ritter}},\ }%
  \bibfield{journal}{%
  \bibinfo {journal} {Phys. Rev.}\ }%
  \textbf{\bibinfo {volume} {C60}},\ \bibinfo {pages} {024901} (\bibinfo {year}
  {1999})%
  \bibAnnoteFile{NoStop}{Voloshin:1999yf}%
\bibitem{Baym:1999up}%
  \BibitemOpen
  \bibfield{author}{%
  \bibinfo {author} {\bibfnamefont{G.}~\bibnamefont{Baym}}\ and\ \bibinfo
  {author} {\bibfnamefont{H.}~\bibnamefont{Heiselberg}},\ }%
  \bibfield{journal}{%
  \bibinfo {journal} {Phys. Lett.}\ }%
  \textbf{\bibinfo {volume} {B469}},\ \bibinfo {pages} {7} (\bibinfo {year}
  {1999})%
  \bibAnnoteFile{NoStop}{Baym:1999up}%
\bibitem{Voloshin:2001ei}%
  \BibitemOpen
  \bibfield{author}{%
  \bibinfo {author} {\bibfnamefont{S.~A.}\ \bibnamefont{Voloshin}} (\bibinfo
  {collaboration} {STAR Collaboration}),\ }%
  \bibfield{journal}{%
  \Doi{10.1063/1.1469997}{\bibinfo {journal} {AIP Conf. Proc.}}\ }%
  \textbf{\bibinfo {volume} {610}},\ \bibinfo {pages} {591} (\bibinfo {year}
  {2001})%
  \bibAnnoteFile{NoStop}{Voloshin:2001ei}%
\bibitem{Korus:2001au}%
  \BibitemOpen
  \bibfield{author}{%
  \bibinfo {author} {\bibfnamefont{R.}~\bibnamefont{Korus}}, \bibinfo {author}
  {\bibfnamefont{S.}~\bibnamefont{Mrowczynski}}, \bibinfo {author}
  {\bibfnamefont{M.}~\bibnamefont{Rybczy\'nski}},\ and\ \bibinfo {author}
  {\bibfnamefont{Z.}~\bibnamefont{Wlodarczyk}},\ }%
  \bibfield{journal}{%
  \bibinfo {journal} {Phys. Rev.}\ }%
  \textbf{\bibinfo {volume} {C64}},\ \bibinfo {pages} {054908} (\bibinfo {year}
  {2001})%
  \bibAnnoteFile{NoStop}{Korus:2001au}%
\bibitem{Gavin:2003cb}%
  \BibitemOpen
  \bibfield{author}{%
  \bibinfo {author} {\bibfnamefont{S.}~\bibnamefont{Gavin}},\ }%
  \bibfield{journal}{%
  \bibinfo {journal} {Phys. Rev. Lett.}\ }%
  \textbf{\bibinfo {volume} {92}},\ \bibinfo {pages} {162301} (\bibinfo {year}
  {2004})%
  \bibAnnoteFile{NoStop}{Gavin:2003cb}%
\bibitem{DiasdeDeus:2003ei}%
  \BibitemOpen
  \bibfield{author}{%
  \bibinfo {author} {\bibfnamefont{J.}~\bibnamefont{Dias~de Deus}}, \bibinfo
  {author} {\bibfnamefont{E.}~\bibnamefont{Ferreiro}}, \bibinfo {author}
  {\bibfnamefont{C.}~\bibnamefont{Pajares}},\ and\ \bibinfo {author}
  {\bibfnamefont{R.}~\bibnamefont{Ugoccioni}},\ }%
  \bibfield{journal}{%
  \Doi{10.1140/epjc/s2005-02127-y}{\bibinfo {journal} {Eur.Phys.J.}}\ }%
  \textbf{\bibinfo {volume} {C40}},\ \bibinfo {pages} {229} (\bibinfo {year}
  {2005})%
  \bibAnnoteFile{NoStop}{DiasdeDeus:2003ei}%
\bibitem{Voloshin:2004th}%
  \BibitemOpen
  \bibfield{author}{%
  \bibinfo {author} {\bibfnamefont{S.~A.}\ \bibnamefont{Voloshin}},\ }%
  \bibfield{journal}{%
  \Doi{10.1016/j.nuclphysa.2004.12.053}{\bibinfo {journal} {Nucl.Phys.}}\ }%
  \textbf{\bibinfo {volume} {A749}},\ \bibinfo {pages} {287} (\bibinfo {year}
  {2005})%
  \bibAnnoteFile{NoStop}{Voloshin:2004th}%
\bibitem{Mrowczynski:2004cg}%
  \BibitemOpen
  \bibfield{author}{%
  \bibinfo {author} {\bibfnamefont{S.}~\bibnamefont{Mrowczynski}}, \bibinfo
  {author} {\bibfnamefont{M.}~\bibnamefont{Rybczy\'nski}},\ and\ \bibinfo
  {author} {\bibfnamefont{Z.}~\bibnamefont{Wlodarczyk}},\ }%
  \bibfield{journal}{%
  \Doi{10.1103/PhysRevC.70.054906}{\bibinfo {journal} {Phys. Rev.}}\ }%
  \textbf{\bibinfo {volume} {C70}},\ \bibinfo {pages} {054906} (\bibinfo {year}
  {2004})%
  \bibAnnoteFile{NoStop}{Mrowczynski:2004cg}%
\bibitem{AbdelAziz:2005wc}%
  \BibitemOpen
  \bibfield{author}{%
  \bibinfo {author} {\bibfnamefont{M.}~\bibnamefont{Abdel-Aziz}}\ and\ \bibinfo
  {author} {\bibfnamefont{S.}~\bibnamefont{Gavin}},\ }%
  \bibfield{journal}{%
  \Doi{10.1016/j.nuclphysa.2006.06.100}{\bibinfo {journal} {Nucl.Phys.}}\ }%
  \textbf{\bibinfo {volume} {A774}},\ \bibinfo {pages} {623} (\bibinfo {year}
  {2006})%
  \bibAnnoteFile{NoStop}{AbdelAziz:2005wc}%
\bibitem{Broniowski:2005ae}%
  \BibitemOpen
  \bibfield{author}{%
  \bibinfo {author} {\bibfnamefont{W.}~\bibnamefont{Broniowski}}, \bibinfo
  {author} {\bibfnamefont{B.}~\bibnamefont{Hiller}}, \bibinfo {author}
  {\bibfnamefont{W.}~\bibnamefont{Florkowski}},\ and\ \bibinfo {author}
  {\bibfnamefont{P.}~\bibnamefont{Bo\.zek}},\ }%
  \bibfield{journal}{%
  \Doi{10.1016/j.physletb.2006.02.056}{\bibinfo {journal} {Phys. Lett.}}\ }%
  \textbf{\bibinfo {volume} {B635}},\ \bibinfo {pages} {290} (\bibinfo {year}
  {2006})%
  \bibAnnoteFile{NoStop}{Broniowski:2005ae}%
\bibitem{Prindle:2006zz}%
  \BibitemOpen
  \bibfield{author}{%
  \bibinfo {author} {\bibfnamefont{D.~J.}\ \bibnamefont{Prindle}}\ and\
  \bibinfo {author} {\bibfnamefont{T.~A.}\ \bibnamefont{Trainor}} (\bibinfo
  {collaboration} {STAR Collaboration}),\ }%
  \bibfield{journal}{%
  \bibinfo {journal} {PoS}\ }%
  \textbf{\bibinfo {volume} {CFRNC2006}},\ \bibinfo {pages} {007} (\bibinfo
  {year} {2006})%
  \bibAnnoteFile{NoStop}{Prindle:2006zz}%
\bibitem{Gavin:2006xd}%
  \BibitemOpen
  \bibfield{author}{%
  \bibinfo {author} {\bibfnamefont{S.}~\bibnamefont{Gavin}}\ and\ \bibinfo
  {author} {\bibfnamefont{M.}~\bibnamefont{Abdel-Aziz}},\ }%
  \bibfield{journal}{%
  \Doi{10.1103/PhysRevLett.97.162302}{\bibinfo {journal} {Phys. Rev. Lett.}}\
  }%
  \textbf{\bibinfo {volume} {97}},\ \bibinfo {pages} {162302} (\bibinfo {year}
  {2006})%
  \bibAnnoteFile{NoStop}{Gavin:2006xd}%
\bibitem{Sharma:2008qr}%
  \BibitemOpen
  \bibfield{author}{%
  \bibinfo {author} {\bibfnamefont{M.}~\bibnamefont{Sharma}}\ and\ \bibinfo
  {author} {\bibfnamefont{C.~A.}\ \bibnamefont{Pruneau}},\ }%
  \bibfield{journal}{%
  \Doi{10.1103/PhysRevC.79.024905}{\bibinfo {journal} {Phys. Rev.}}\ }%
  \textbf{\bibinfo {volume} {C79}},\ \bibinfo {pages} {024905} (\bibinfo {year}
  {2009})%
  \bibAnnoteFile{NoStop}{Sharma:2008qr}%
\bibitem{Mrowczynski:2009wk}%
  \BibitemOpen
  \bibfield{author}{%
  \bibinfo {author} {\bibfnamefont{S.}~\bibnamefont{Mrowczynski}},\ }%
  \bibfield{journal}{%
  \bibinfo {journal} {Acta Phys.Polon.}\ }%
  \textbf{\bibinfo {volume} {B40}},\ \bibinfo {pages} {1053} (\bibinfo {year}
  {2009})%
  \bibAnnoteFile{NoStop}{Mrowczynski:2009wk}%
\bibitem{Hama:2009pk}%
  \BibitemOpen
  \bibfield{author}{%
  \bibinfo {author} {\bibfnamefont{Y.}~\bibnamefont{Hama}}, \bibinfo {author}
  {\bibfnamefont{R.~P.~G.}\ \bibnamefont{Andrade}}, \bibinfo {author}
  {\bibfnamefont{F.}~\bibnamefont{Grassi}}, \bibinfo {author}
  {\bibfnamefont{W.~L.}\ \bibnamefont{Qian}},\ and\ \bibinfo {author}
  {\bibfnamefont{T.}~\bibnamefont{Kodama}},\ }%
  \bibfield{journal}{%
  \bibinfo {journal} {Acta Phys. Polon.}\ }%
  \textbf{\bibinfo {volume} {B40}},\ \bibinfo {pages} {931} (\bibinfo {year}
  {2009})%
  \bibAnnoteFile{NoStop}{Hama:2009pk}%
\bibitem{Trainor:2015swa}%
  \BibitemOpen
  \bibfield{author}{%
  \bibinfo {author} {\bibfnamefont{T.~A.}\ \bibnamefont{Trainor}},\ }%
  \bibfield{journal}{%
  \Doi{10.1103/PhysRevC.92.024915}{\bibinfo {journal} {Phys. Rev.}}\ }%
  \textbf{\bibinfo {volume} {C92}},\ \bibinfo {pages} {024915} (\bibinfo {year}
  {2015})%
  \bibAnnoteFile{NoStop}{Trainor:2015swa}%
\bibitem{Liu:2016apq}%
  \BibitemOpen
  \bibfield{author}{%
  \bibinfo {author} {\bibfnamefont{Q.}~\bibnamefont{Liu}}\ and\ \bibinfo
  {author} {\bibfnamefont{W.-Q.}\ \bibnamefont{Zhao}}}%
   (\bibinfo {year} {2016}),\
  \Eprint{http://arxiv.org/abs/1611.02532}{arXiv:1611.02532 [hep-ph]}%
  \bibAnnoteFile{NoStop}{Liu:2016apq}%
\bibitem{Adams:2003uw}%
  \BibitemOpen
  \bibfield{author}{%
  \bibinfo {author} {\bibfnamefont{J.}~\bibnamefont{Adams}} \emph{et~al.}
  (\bibinfo {collaboration} {STAR Collaboration}),\ }%
  \bibfield{journal}{%
  \Doi{10.1103/PhysRevC.71.064906}{\bibinfo {journal} {Phys. Rev.}}\ }%
  \textbf{\bibinfo {volume} {C71}},\ \bibinfo {pages} {064906} (\bibinfo {year}
  {2005})%
  \bibAnnoteFile{NoStop}{Adams:2003uw}%
\bibitem{Adamova:2003pz}%
  \BibitemOpen
  \bibfield{author}{%
  \bibinfo {author} {\bibfnamefont{D.}~\bibnamefont{Adamova}} \emph{et~al.}
  (\bibinfo {collaboration} {CERES Collaboration}),\ }%
  \bibfield{journal}{%
  \Doi{10.1016/j.nuclphysa.2003.07.018}{\bibinfo {journal} {Nucl.Phys.}}\ }%
  \textbf{\bibinfo {volume} {A727}},\ \bibinfo {pages} {97} (\bibinfo {year}
  {2003})%
  \bibAnnoteFile{NoStop}{Adamova:2003pz}%
\bibitem{Adler:2003xq}%
  \BibitemOpen
  \bibfield{author}{%
  \bibinfo {author} {\bibfnamefont{S.}~\bibnamefont{Adler}} \emph{et~al.}
  (\bibinfo {collaboration} {PHENIX Collaboration}),\ }%
  \bibfield{journal}{%
  \Doi{10.1103/PhysRevLett.93.092301}{\bibinfo {journal} {Phys. Rev. Lett.}}\
  }%
  \textbf{\bibinfo {volume} {93}},\ \bibinfo {pages} {092301} (\bibinfo {year}
  {2004})%
  \bibAnnoteFile{NoStop}{Adler:2003xq}%
\bibitem{Anticic:2003fd}%
  \BibitemOpen
  \bibfield{author}{%
  \bibinfo {author} {\bibfnamefont{T.}~\bibnamefont{Anticic}} \emph{et~al.}
  (\bibinfo {collaboration} {NA49 Collaboration}),\ }%
  \bibfield{journal}{%
  \Doi{10.1103/PhysRevC.70.034902}{\bibinfo {journal} {Phys. Rev.}}\ }%
  \textbf{\bibinfo {volume} {C70}},\ \bibinfo {pages} {034902} (\bibinfo {year}
  {2004})%
  \bibAnnoteFile{NoStop}{Anticic:2003fd}%
\bibitem{Adams:2004gp}%
  \BibitemOpen
  \bibfield{author}{%
  \bibinfo {author} {\bibfnamefont{J.}~\bibnamefont{Adams}} \emph{et~al.}
  (\bibinfo {collaboration} {STAR Collaboration}),\ }%
  \bibfield{journal}{%
  \Doi{10.1088/0954-3899/34/5/002}{\bibinfo {journal} {J.Phys.G}}\ }%
  \textbf{\bibinfo {volume} {G34}},\ \bibinfo {pages} {799} (\bibinfo {year}
  {2007})%
  \bibAnnoteFile{NoStop}{Adams:2004gp}%
\bibitem{Adams:2005ka}%
  \BibitemOpen
  \bibfield{author}{%
  \bibinfo {author} {\bibfnamefont{J.}~\bibnamefont{Adams}} \emph{et~al.}
  (\bibinfo {collaboration} {STAR Collaboration}),\ }%
  \bibfield{journal}{%
  \Doi{10.1103/PhysRevC.72.044902}{\bibinfo {journal} {Phys. Rev.}}\ }%
  \textbf{\bibinfo {volume} {C72}},\ \bibinfo {pages} {044902} (\bibinfo {year}
  {2005})%
  \bibAnnoteFile{NoStop}{Adams:2005ka}%
\bibitem{Adams:2005aw}%
  \BibitemOpen
  \bibfield{author}{%
  \bibinfo {author} {\bibfnamefont{J.}~\bibnamefont{Adams}} \emph{et~al.}
  (\bibinfo {collaboration} {STAR Collaboration}),\ }%
  \bibfield{journal}{%
  \Doi{10.1088/0954-3899/32/6/L02}{\bibinfo {journal} {J.Phys.G}}\ }%
  \textbf{\bibinfo {volume} {G32}},\ \bibinfo {pages} {L37} (\bibinfo {year}
  {2006})%
  \bibAnnoteFile{NoStop}{Adams:2005aw}%
\bibitem{Adams:2006sg}%
  \BibitemOpen
  \bibfield{author}{%
  \bibinfo {author} {\bibfnamefont{J.}~\bibnamefont{Adams}} \emph{et~al.}
  (\bibinfo {collaboration} {STAR Collaboration}),\ }%
  \bibfield{journal}{%
  \Doi{10.1088/0954-3899/34/3/004}{\bibinfo {journal} {J.Phys.G}}\ }%
  \textbf{\bibinfo {volume} {G34}},\ \bibinfo {pages} {451} (\bibinfo {year}
  {2007})%
  \bibAnnoteFile{NoStop}{Adams:2006sg}%
\bibitem{Agakishiev:2011fs}%
  \BibitemOpen
  \bibfield{author}{%
  \bibinfo {author} {\bibfnamefont{H.}~\bibnamefont{Agakishiev}} \emph{et~al.}
  (\bibinfo {collaboration} {STAR Collaboration}),\ }%
  \bibfield{journal}{%
  \Doi{10.1016/j.physletb.2011.09.075}{\bibinfo {journal} {Phys. Lett.}}\ }%
  \textbf{\bibinfo {volume} {B704}},\ \bibinfo {pages} {467} (\bibinfo {year}
  {2011})%
  \bibAnnoteFile{NoStop}{Agakishiev:2011fs}%
\bibitem{Glauber:1959aa}%
  \BibitemOpen
  \bibfield{author}{%
  \bibinfo {author} {\bibfnamefont{R.~J.}\ \bibnamefont{Glauber}}\ }%
  \bibinfo {note} {~in {\it Lectures in Theoretical Physics} W.~E. Brittin and
  L.~G. Dunham eds., (Interscience, New York, 1959) Vol. 1, p. 315}%
  \bibAnnoteFile{NoStop}{Glauber:1959aa}%
\bibitem{Czyz:1969jg}%
  \BibitemOpen
  \bibfield{author}{%
  \bibinfo {author} {\bibfnamefont{W.}~\bibnamefont{Czy\.z}}\ and\ \bibinfo
  {author} {\bibfnamefont{L.~C.}\ \bibnamefont{Maximon}},\ }%
  \bibfield{journal}{%
  \Doi{10.1016/0003-4916(69)90321-2}{\bibinfo {journal} {Annals Phys.}}\ }%
  \textbf{\bibinfo {volume} {52}},\ \bibinfo {pages} {59} (\bibinfo {year}
  {1969})%
  \bibAnnoteFile{NoStop}{Czyz:1969jg}%
\bibitem{Bialas:1976ed}%
  \BibitemOpen
  \bibfield{author}{%
  \bibinfo {author} {\bibfnamefont{A.}~\bibnamefont{Bia\l{}as}}, \bibinfo
  {author} {\bibfnamefont{M.}~\bibnamefont{B\l{}eszy\'nski}},\ and\ \bibinfo
  {author} {\bibfnamefont{W.}~\bibnamefont{Czy\.z}},\ }%
  \bibfield{journal}{%
  \bibinfo {journal} {Nucl. Phys.}\ }%
  \textbf{\bibinfo {volume} {B111}},\ \bibinfo {pages} {461} (\bibinfo {year}
  {1976})%
  \bibAnnoteFile{NoStop}{Bialas:1976ed}%
\bibitem{Bialas:2008zza}%
  \BibitemOpen
  \bibfield{author}{%
  \bibinfo {author} {\bibfnamefont{A.}~\bibnamefont{Bia\l{}as}},\ }%
  \bibfield{journal}{%
  \Doi{10.1088/0954-3899/35/4/044053}{\bibinfo {journal} {J. Phys.}}\ }%
  \textbf{\bibinfo {volume} {G35}},\ \bibinfo {pages} {044053} (\bibinfo {year}
  {2008})%
  \bibAnnoteFile{NoStop}{Bialas:2008zza}%
\bibitem{Eremin:2003qn}%
  \BibitemOpen
  \bibfield{author}{%
  \bibinfo {author} {\bibfnamefont{S.}~\bibnamefont{Eremin}}\ and\ \bibinfo
  {author} {\bibfnamefont{S.}~\bibnamefont{Voloshin}},\ }%
  \bibfield{journal}{%
  \Doi{10.1103/PhysRevC.67.064905}{\bibinfo {journal} {Phys. Rev.}}\ }%
  \textbf{\bibinfo {volume} {C67}},\ \bibinfo {pages} {064905} (\bibinfo {year}
  {2003})%
  \bibAnnoteFile{NoStop}{Eremin:2003qn}%
\bibitem{KumarNetrakanti:2004ym}%
  \BibitemOpen
  \bibfield{author}{%
  \bibinfo {author} {\bibfnamefont{P.}~\bibnamefont{Kumar~Netrakanti}}\ and\
  \bibinfo {author} {\bibfnamefont{B.}~\bibnamefont{Mohanty}},\ }%
  \bibfield{journal}{%
  \Doi{10.1103/PhysRevC.70.027901}{\bibinfo {journal} {Phys. Rev.}}\ }%
  \textbf{\bibinfo {volume} {C70}},\ \bibinfo {pages} {027901} (\bibinfo {year}
  {2004})%
  \bibAnnoteFile{NoStop}{KumarNetrakanti:2004ym}%
\bibitem{Agakishiev:2011eq}%
  \BibitemOpen
  \bibfield{author}{%
  \bibinfo {author} {\bibfnamefont{G.}~\bibnamefont{Agakishiev}} \emph{et~al.}
  (\bibinfo {collaboration} {STAR}),\ }%
  \bibfield{journal}{%
  \Doi{10.1103/PhysRevC.86.014904}{\bibinfo {journal} {Phys. Rev.}}\ }%
  \textbf{\bibinfo {volume} {C86}},\ \bibinfo {pages} {014904} (\bibinfo {year}
  {2012})%
  \bibAnnoteFile{NoStop}{Agakishiev:2011eq}%
\bibitem{Adler:2013aqf}%
  \BibitemOpen
  \bibfield{author}{%
  \bibinfo {author} {\bibfnamefont{S.~S.}\ \bibnamefont{Adler}} \emph{et~al.}
  (\bibinfo {collaboration} {PHENIX}),\ }%
  \bibfield{journal}{%
  \Doi{10.1103/PhysRevC.89.044905}{\bibinfo {journal} {Phys. Rev.}}\ }%
  \textbf{\bibinfo {volume} {C89}},\ \bibinfo {pages} {044905} (\bibinfo {year}
  {2014})%
  \bibAnnoteFile{NoStop}{Adler:2013aqf}%
\bibitem{Loizides:2014vua}%
  \BibitemOpen
  \bibfield{author}{%
  \bibinfo {author} {\bibfnamefont{C.}~\bibnamefont{Loizides}}, \bibinfo
  {author} {\bibfnamefont{J.}~\bibnamefont{Nagle}},\ and\ \bibinfo {author}
  {\bibfnamefont{P.}~\bibnamefont{Steinberg}},\ }%
  \bibfield{journal}{%
  \Doi{10.1016/j.softx.2015.05.001}{\bibinfo {journal} {SoftwareX}}\ }%
  \textbf{\bibinfo {volume} {1-2}},\ \bibinfo {pages} {13} (\bibinfo {year}
  {2015})%
  \bibAnnoteFile{NoStop}{Loizides:2014vua}%
\bibitem{Adare:2015bua}%
  \BibitemOpen
  \bibfield{author}{%
  \bibinfo {author} {\bibfnamefont{A.}~\bibnamefont{Adare}} \emph{et~al.}
  (\bibinfo {collaboration} {PHENIX}),\ }%
  \bibfield{journal}{%
  \Doi{10.1103/PhysRevC.93.024901}{\bibinfo {journal} {Phys. Rev.}}\ }%
  \textbf{\bibinfo {volume} {C93}},\ \bibinfo {pages} {024901} (\bibinfo {year}
  {2016})%
  \bibAnnoteFile{NoStop}{Adare:2015bua}%
\bibitem{Zheng:2016nxx}%
  \BibitemOpen
  \bibfield{author}{%
  \bibinfo {author} {\bibfnamefont{L.}~\bibnamefont{Zheng}}\ and\ \bibinfo
  {author} {\bibfnamefont{Z.}~\bibnamefont{Yin}},\ }%
  \bibfield{journal}{%
  \Doi{10.1140/epja/i2016-16045-x}{\bibinfo {journal} {Eur. Phys. J.}}\ }%
  \textbf{\bibinfo {volume} {A52}},\ \bibinfo {pages} {45} (\bibinfo {year}
  {2016})%
  \bibAnnoteFile{NoStop}{Zheng:2016nxx}%
\bibitem{Mitchell:2016jio}%
  \BibitemOpen
  \bibfield{author}{%
  \bibinfo {author} {\bibfnamefont{J.~T.}\ \bibnamefont{Mitchell}}, \bibinfo
  {author} {\bibfnamefont{D.~V.}\ \bibnamefont{Perepelitsa}}, \bibinfo {author}
  {\bibfnamefont{M.~J.}\ \bibnamefont{Tannenbaum}},\ and\ \bibinfo {author}
  {\bibfnamefont{P.~W.}\ \bibnamefont{Stankus}}}%
   (\bibinfo {year} {2016}),\
  \Eprint{http://arxiv.org/abs/1603.08836}{arXiv:1603.08836 [nucl-ex]}%
  \bibAnnoteFile{NoStop}{Mitchell:2016jio}%
\bibitem{Loizides:2016djv}%
  \BibitemOpen
  \bibfield{author}{%
  \bibinfo {author} {\bibfnamefont{C.}~\bibnamefont{Loizides}},\ }%
  \bibfield{journal}{%
  \Doi{10.1103/PhysRevC.94.024914}{\bibinfo {journal} {Phys. Rev.}}\ }%
  \textbf{\bibinfo {volume} {C94}},\ \bibinfo {pages} {024914} (\bibinfo {year}
  {2016})%
  \bibAnnoteFile{NoStop}{Loizides:2016djv}%
\bibitem{Bialas:2006kw}%
  \BibitemOpen
  \bibfield{author}{%
  \bibinfo {author} {\bibfnamefont{A.}~\bibnamefont{Bia\l{}as}}\ and\ \bibinfo
  {author} {\bibfnamefont{A.}~\bibnamefont{Bzdak}},\ }%
  \bibfield{journal}{%
  \bibinfo {journal} {Phys. Lett.}\ }%
  \textbf{\bibinfo {volume} {B649}},\ \bibinfo {pages} {263} (\bibinfo {year}
  {2007})%
  \bibAnnoteFile{NoStop}{Bialas:2006kw}%
\bibitem{Bialas:2007eg}%
  \BibitemOpen
  \bibfield{author}{%
  \bibinfo {author} {\bibfnamefont{A.}~\bibnamefont{Bia\l{}as}}\ and\ \bibinfo
  {author} {\bibfnamefont{A.}~\bibnamefont{Bzdak}},\ }%
  \bibfield{journal}{%
  \Doi{10.1103/PhysRevC.77.034908}{\bibinfo {journal} {Phys. Rev.}}\ }%
  \textbf{\bibinfo {volume} {C77}},\ \bibinfo {pages} {034908} (\bibinfo {year}
  {2008})%
  \bibAnnoteFile{NoStop}{Bialas:2007eg}%
\bibitem{Kharzeev:2000ph}%
  \BibitemOpen
  \bibfield{author}{%
  \bibinfo {author} {\bibfnamefont{D.}~\bibnamefont{Kharzeev}}\ and\ \bibinfo
  {author} {\bibfnamefont{M.}~\bibnamefont{Nardi}},\ }%
  \bibfield{journal}{%
  \Doi{10.1016/S0370-2693(01)00457-9}{\bibinfo {journal} {Phys. Lett.}}\ }%
  \textbf{\bibinfo {volume} {B507}},\ \bibinfo {pages} {121} (\bibinfo {year}
  {2001})%
  \bibAnnoteFile{NoStop}{Kharzeev:2000ph}%
\bibitem{Back:2001xy}%
  \BibitemOpen
  \bibfield{author}{%
  \bibinfo {author} {\bibfnamefont{B.~B.}\ \bibnamefont{Back}} \emph{et~al.}
  (\bibinfo {collaboration} {PHOBOS}),\ }%
  \bibfield{journal}{%
  \bibinfo {journal} {Phys. Rev.}\ }%
  \textbf{\bibinfo {volume} {C65}},\ \bibinfo {pages} {031901} (\bibinfo {year}
  {2002})%
  \bibAnnoteFile{NoStop}{Back:2001xy}%
\bibitem{Bozek:2012qs}%
  \BibitemOpen
  \bibfield{author}{%
  \bibinfo {author} {\bibfnamefont{P.}~\bibnamefont{Bo\.zek}}\ and\ \bibinfo
  {author} {\bibfnamefont{I.}~\bibnamefont{Wyskiel-Piekarska}},\ }%
  \bibfield{journal}{%
  \bibinfo {journal} {Phys. Rev.}\ }%
  \textbf{\bibinfo {volume} {C85}},\ \bibinfo {pages} {064915} (\bibinfo {year}
  {2012})%
  \bibAnnoteFile{NoStop}{Bozek:2012qs}%
\bibitem{Albacete:2016pmp}%
  \BibitemOpen
  \bibfield{author}{%
  \bibinfo {author} {\bibfnamefont{J.~L.}\ \bibnamefont{Albacete}}\ and\
  \bibinfo {author} {\bibfnamefont{A.}~\bibnamefont{Soto-Ontoso}}}%
   (\bibinfo {year} {2016}),\
  \Eprint{http://arxiv.org/abs/1605.09176}{arXiv:1605.09176 [hep-ph]}%
  \bibAnnoteFile{NoStop}{Albacete:2016pmp}%
\bibitem{Kovner:2002xa}%
  \BibitemOpen
  \bibfield{author}{%
  \bibinfo {author} {\bibfnamefont{A.}~\bibnamefont{Kovner}}\ and\ \bibinfo
  {author} {\bibfnamefont{U.~A.}\ \bibnamefont{Wiedemann}},\ }%
  \bibfield{journal}{%
  \Doi{10.1103/PhysRevD.66.034031}{\bibinfo {journal} {Phys. Rev.}}\ }%
  \textbf{\bibinfo {volume} {D66}},\ \bibinfo {pages} {034031} (\bibinfo {year}
  {2002})%
  \bibAnnoteFile{NoStop}{Kovner:2002xa}%
\bibitem{Broniowski:2007nz}%
  \BibitemOpen
  \bibfield{author}{%
  \bibinfo {author} {\bibfnamefont{W.}~\bibnamefont{Broniowski}}, \bibinfo
  {author} {\bibfnamefont{M.}~\bibnamefont{Rybczy\'nski}},\ and\ \bibinfo
  {author} {\bibfnamefont{P.}~\bibnamefont{Bo\.zek}},\ }%
  \bibfield{journal}{%
  \Doi{10.1016/j.cpc.2008.07.016}{\bibinfo {journal} {Comput. Phys. Commun.}}\
  }%
  \textbf{\bibinfo {volume} {180}},\ \bibinfo {pages} {69} (\bibinfo {year}
  {2009})%
  \bibAnnoteFile{NoStop}{Broniowski:2007nz}%
\bibitem{Rybczynski:2013yba}%
  \BibitemOpen
  \bibfield{author}{%
  \bibinfo {author} {\bibfnamefont{M.}~\bibnamefont{Rybczy\'nski}}, \bibinfo
  {author} {\bibfnamefont{G.}~\bibnamefont{Stefanek}}, \bibinfo {author}
  {\bibfnamefont{W.}~\bibnamefont{Broniowski}},\ and\ \bibinfo {author}
  {\bibfnamefont{P.}~\bibnamefont{Bo\.zek}},\ }%
  \bibfield{journal}{%
  \Doi{10.1016/j.cpc.2014.02.016}{\bibinfo {journal} {Comput. Phys. Commun.}}\
  }%
  \textbf{\bibinfo {volume} {185}},\ \bibinfo {pages} {1759} (\bibinfo {year}
  {2014})%
  \bibAnnoteFile{NoStop}{Rybczynski:2013yba}%
\bibitem{Bhalerao:2005mm}%
  \BibitemOpen
  \bibfield{author}{%
  \bibinfo {author} {\bibfnamefont{R.~S.}\ \bibnamefont{Bhalerao}}, \bibinfo
  {author} {\bibfnamefont{J.-P.}\ \bibnamefont{Blaizot}}, \bibinfo {author}
  {\bibfnamefont{N.}~\bibnamefont{Borghini}},\ and\ \bibinfo {author}
  {\bibfnamefont{J.-Y.}\ \bibnamefont{Ollitrault}},\ }%
  \bibfield{journal}{%
  \Doi{10.1016/j.physletb.2005.08.131}{\bibinfo {journal} {Phys. Lett.}}\ }%
  \textbf{\bibinfo {volume} {B627}},\ \bibinfo {pages} {49} (\bibinfo {year}
  {2005})%
  \bibAnnoteFile{NoStop}{Bhalerao:2005mm}%
\bibitem{Schenke:2010rr}%
  \BibitemOpen
  \bibfield{author}{%
  \bibinfo {author} {\bibfnamefont{B.}~\bibnamefont{Schenke}}, \bibinfo
  {author} {\bibfnamefont{S.}~\bibnamefont{Jeon}},\ and\ \bibinfo {author}
  {\bibfnamefont{C.}~\bibnamefont{Gale}},\ }%
  \bibfield{journal}{%
  \Doi{10.1103/PhysRevLett.106.042301}{\bibinfo {journal} {Phys. Rev. Lett.}}\
  }%
  \textbf{\bibinfo {volume} {106}},\ \bibinfo {pages} {042301} (\bibinfo {year}
  {2011})%
  \bibAnnoteFile{NoStop}{Schenke:2010rr}%
\bibitem{Borsanyi:2010cj}%
  \BibitemOpen
  \bibfield{author}{%
  \bibinfo {author} {\bibfnamefont{S.}~\bibnamefont{Borsanyi}} \emph{et~al.},\
  }%
  \bibfield{journal}{%
  \Doi{10.1007/JHEP11(2010)077}{\bibinfo {journal} {JHEP}}\ }%
  \textbf{\bibinfo {volume} {11}},\ \bibinfo {pages} {077} (\bibinfo {year}
  {2010})%
  \bibAnnoteFile{NoStop}{Borsanyi:2010cj}%
\bibitem{Chojnacki:2007jc}%
  \BibitemOpen
  \bibfield{author}{%
  \bibinfo {author} {\bibfnamefont{M.}~\bibnamefont{Chojnacki}}\ and\ \bibinfo
  {author} {\bibfnamefont{W.}~\bibnamefont{Florkowski}},\ }%
  \bibfield{journal}{%
  \bibinfo {journal} {Acta Phys. Polon.}\ }%
  \textbf{\bibinfo {volume} {B38}},\ \bibinfo {pages} {3249} (\bibinfo {year}
  {2007})%
  \bibAnnoteFile{NoStop}{Chojnacki:2007jc}%
\bibitem{Bozek:2011ua}%
  \BibitemOpen
  \bibfield{author}{%
  \bibinfo {author} {\bibfnamefont{P.}~\bibnamefont{Bo\.zek}},\ }%
  \bibfield{journal}{%
  \Doi{10.1103/PhysRevC.85.034901}{\bibinfo {journal} {Phys. Rev.}}\ }%
  \textbf{\bibinfo {volume} {C85}},\ \bibinfo {pages} {034901} (\bibinfo {year}
  {2012})%
  \bibAnnoteFile{NoStop}{Bozek:2011ua}%
\bibitem{Cooper:1974mv}%
  \BibitemOpen
  \bibfield{author}{%
  \bibinfo {author} {\bibfnamefont{F.}~\bibnamefont{Cooper}}\ and\ \bibinfo
  {author} {\bibfnamefont{G.}~\bibnamefont{Frye}},\ }%
  \bibfield{journal}{%
  \Doi{10.1103/PhysRevD.10.186}{\bibinfo {journal} {Phys. Rev.}}\ }%
  \textbf{\bibinfo {volume} {D10}},\ \bibinfo {pages} {186} (\bibinfo {year}
  {1974})%
  \bibAnnoteFile{NoStop}{Cooper:1974mv}%
\bibitem{Kisiel:2005hn}%
  \BibitemOpen
  \bibfield{author}{%
  \bibinfo {author} {\bibfnamefont{A.}~\bibnamefont{Kisiel}}, \bibinfo {author}
  {\bibfnamefont{T.}~\bibnamefont{{Ta\l{}u\'c}}}, \bibinfo {author}
  {\bibfnamefont{W.}~\bibnamefont{Broniowski}},\ and\ \bibinfo {author}
  {\bibfnamefont{W.}~\bibnamefont{Florkowski}},\ }%
  \bibfield{journal}{%
  \bibinfo {journal} {Comput. Phys. Commun.}\ }%
  \textbf{\bibinfo {volume} {174}},\ \bibinfo {pages} {669} (\bibinfo {year}
  {2006})%
  \bibAnnoteFile{NoStop}{Kisiel:2005hn}%
\bibitem{Chojnacki:2011hb}%
  \BibitemOpen
  \bibfield{author}{%
  \bibinfo {author} {\bibfnamefont{M.}~\bibnamefont{Chojnacki}}, \bibinfo
  {author} {\bibfnamefont{A.}~\bibnamefont{Kisiel}}, \bibinfo {author}
  {\bibfnamefont{W.}~\bibnamefont{Florkowski}},\ and\ \bibinfo {author}
  {\bibfnamefont{W.}~\bibnamefont{Broniowski}},\ }%
  \bibfield{journal}{%
  \Doi{10.1016/j.cpc.2011.11.018}{\bibinfo {journal} {Comput. Phys. Commun.}}\
  }%
  \textbf{\bibinfo {volume} {183}},\ \bibinfo {pages} {746} (\bibinfo {year}
  {2012})%
  \bibAnnoteFile{NoStop}{Chojnacki:2011hb}%
\bibitem{Gale:2013da}%
  \BibitemOpen
  \bibfield{author}{%
  \bibinfo {author} {\bibfnamefont{C.}~\bibnamefont{Gale}}, \bibinfo {author}
  {\bibfnamefont{S.}~\bibnamefont{Jeon}},\ and\ \bibinfo {author}
  {\bibfnamefont{B.}~\bibnamefont{Schenke}},\ }%
  \bibfield{journal}{%
  \Doi{10.1142/S0217751X13400113}{\bibinfo {journal} {Int.J.Mod.Phys.}}\ }%
  \textbf{\bibinfo {volume} {A28}},\ \bibinfo {pages} {1340011} (\bibinfo
  {year} {2013})%
  \bibAnnoteFile{NoStop}{Gale:2013da}%
\bibitem{Heinz:2013th}%
  \BibitemOpen
  \bibfield{author}{%
  \bibinfo {author} {\bibfnamefont{U.}~\bibnamefont{Heinz}}\ and\ \bibinfo
  {author} {\bibfnamefont{R.}~\bibnamefont{Snellings}},\ }%
  \bibfield{journal}{%
  \Doi{10.1146/annurev-nucl-102212-170540}{\bibinfo {journal}
  {Ann.Rev.Nucl.Part.Sci.}}\ }%
  \textbf{\bibinfo {volume} {63}},\ \bibinfo {pages} {123} (\bibinfo {year}
  {2013})%
  \bibAnnoteFile{NoStop}{Heinz:2013th}%
\bibitem{Jeon:2015dfa}%
  \BibitemOpen
  \bibfield{author}{%
  \bibinfo {author} {\bibfnamefont{S.}~\bibnamefont{Jeon}}\ and\ \bibinfo
  {author} {\bibfnamefont{U.}~\bibnamefont{Heinz}},\ }%
  \bibfield{journal}{%
  \Doi{10.1142/S0218301315300106}{\bibinfo {journal} {Int. J. Mod. Phys.}}\ }%
  \textbf{\bibinfo {volume} {E24}},\ \bibinfo {pages} {1530010} (\bibinfo
  {year} {2015})%
  \bibAnnoteFile{NoStop}{Jeon:2015dfa}%
\bibitem{Bozek:2016jhf}%
  \BibitemOpen
  \bibfield{author}{%
  \bibinfo {author} {\bibfnamefont{P.}~\bibnamefont{Bo{\.z}ek}},\ }%
  \bibfield{journal}{%
  \Doi{10.1016/j.nuclphysa.2016.02.029}{\bibinfo {journal} {Nucl. Phys.}}\ }%
  \textbf{\bibinfo {volume} {A956}},\ \bibinfo {pages} {208} (\bibinfo {year}
  {2016})%
  \bibAnnoteFile{NoStop}{Bozek:2016jhf}%
\bibitem{Mazeliauskas:2015efa}%
  \BibitemOpen
  \bibfield{author}{%
  \bibinfo {author} {\bibfnamefont{A.}~\bibnamefont{Mazeliauskas}}\ and\
  \bibinfo {author} {\bibfnamefont{D.}~\bibnamefont{Teaney}},\ }%
  \bibfield{journal}{%
  \Doi{10.1103/PhysRevC.93.024913}{\bibinfo {journal} {Phys. Rev.}}\ }%
  \textbf{\bibinfo {volume} {C93}},\ \bibinfo {pages} {024913} (\bibinfo {year}
  {2016})%
  \bibAnnoteFile{NoStop}{Mazeliauskas:2015efa}%
\bibitem{Bozek:2012en}%
  \BibitemOpen
  \bibfield{author}{%
  \bibinfo {author} {\bibfnamefont{P.}~\bibnamefont{Bo\.zek}}\ and\ \bibinfo
  {author} {\bibfnamefont{W.}~\bibnamefont{Broniowski}},\ }%
  \bibfield{journal}{%
  \Doi{10.1103/PhysRevLett.109.062301}{\bibinfo {journal} {Phys. Rev. Lett.}}\
  }%
  \textbf{\bibinfo {volume} {109}},\ \bibinfo {pages} {062301} (\bibinfo {year}
  {2012})%
  \bibAnnoteFile{NoStop}{Bozek:2012en}%
\end{thebibliography}%

\end{document}